\documentclass[twocolumn,tighten]{aastex63}
\usepackage{subcaption}
\usepackage{amsmath}
\usepackage{amssymb}
\usepackage{newtxtext,newtxmath}
\usepackage{threeparttable}
\usepackage{ulem}
\usepackage[figuresright]{rotating}
\usepackage{graphicx}
\usepackage{booktabs}

\begin{document}

\title{Star Proper Motions Based on Two-epoch Observations from the SDSS and DESI Imaging Surveys}

\correspondingauthor{Hu Zou}
\email{zouhu@nao.cas.cn}

\author[0009-0004-2243-8289]{Yun-Ao Xiao}
\affiliation{Key Laboratory of Optical Astronomy, National Astronomical Observatories, Chinese Academy of Sciences, Beijing 100101, P.R. China}
\affiliation{University of Chinese Academy of Sciences, Beijing 100039, P.R. China}
\author[0000-0002-6684-3997]{Hu Zou}
\affiliation{Key Laboratory of Optical Astronomy, National Astronomical Observatories, Chinese Academy of Sciences, Beijing 100101, P.R. China}
\affiliation{University of Chinese Academy of Sciences, Beijing 100039, P.R. China}
\author{Xin Xu}
\affiliation{Kapteyn Astronomical Institute, University of Groningen, PO BOX 800, 9700 AV Groningen, The Netherlands}
\author{Lu Feng}
\affiliation{Key Laboratory of Optical Astronomy, National Astronomical Observatories, Chinese Academy of Sciences, Beijing 100101, P.R. China}
\author{Wei-Jian Guo}
\affiliation{Key Laboratory of Optical Astronomy, National Astronomical Observatories, Chinese Academy of Sciences, Beijing 100101, P.R. China}
\author{Wen-Xiong Li}
\affiliation{Key Laboratory of Optical Astronomy, National Astronomical Observatories, Chinese Academy of Sciences, Beijing 100101, P.R. China}
\author{Shu-Fei Liu}
\affiliation{Key Laboratory of Optical Astronomy, National Astronomical Observatories, Chinese Academy of Sciences, Beijing 100101, P.R. China}
\affiliation{University of Chinese Academy of Sciences, Beijing 100039, P.R. China}
\author{Zhi-Xia Shen}
\affiliation{Key Laboratory of Optical Astronomy, National Astronomical Observatories, Chinese Academy of Sciences, Beijing 100101, P.R. China}
\author{Gaurav Singh}
\affiliation{Key Laboratory of Optical Astronomy, National Astronomical Observatories, Chinese Academy of Sciences, Beijing 100101, P.R. China}
\author{Ji-Peng Sui}
\affiliation{Key Laboratory of Optical Astronomy, National Astronomical Observatories, Chinese Academy of Sciences, Beijing 100101, P.R. China}
\affiliation{University of Chinese Academy of Sciences, Beijing 100039, P.R. China}
\author{Jia-Li Wang}
\affiliation{Key Laboratory of Optical Astronomy, National Astronomical Observatories, Chinese Academy of Sciences, Beijing 100101, P.R. China}
\author{Sui-Jian Xue}
\affiliation{Key Laboratory of Optical Astronomy, National Astronomical Observatories, Chinese Academy of Sciences, Beijing 100101, P.R. China}

\begin{abstract}
In this study, we present the construction of a new proper motion catalog utilizing the photometric data from the Sloan Digital Sky Survey (SDSS) and Dark Energy Spectroscopic Instrument (DESI) imaging surveys, with a median time baseline of about 13 years. To mitigate systematic errors, the DESI galaxy positions are employed to establish a reference frame and to correct the position-, magnitude-, and color-dependent discrepancies between SDSS and DESI imaging datasets. Spanning 12,589 square degrees, the catalog encompasses about 223.7 million non-Gaia objects down to $m_r \sim$ 23. Based on 734k quasars, the assessment of the global systematic errors in the DESI-SDSS proper motion catalog yields values of 0.06 mas yr$^{-1}$ for $\mu_{\alpha_{*}}$ and 0.12 mas yr$^{-1}$ for $\mu_{\delta}$. The catalog exhibits precision surpassing 3.4 mas yr$^{-1}$, albeit varying with position, color, and magnitude. An additional evaluation employing approximately 2,644 distant star samples yields an overall precision of approximately 2.5 and 2.9 mas yr$^{-1}$ for $\mu_{\alpha_{*}}$ and $\mu_{\delta}$, respectively. Further comparisons with proper motions from SDSS Stripe 82 reveal a strong consistency between the two datasets. As a practical application, we utilize fainter non-Gaia objects in our catalog to update the proper motions of 15 star clusters. The resulting proper motions for these clusters exhibit excellent consistency with those derived from Gaia data. Our proper motion measurements, characterized by a deeper limiting magnitude, stand as a valuable complement to the Gaia dataset.
\end{abstract}

\keywords{Astrometry – Astronomical databases: miscellaneous – Proper Motions}

\section{Introduction}
\label{sec:1}

The vast repository of knowledge derived from the Milky Way is a cornerstone of modern astrophysics. For over a century, the scientific community has dedicated substantial effort to elucidating the intricate structure and dynamic behavior of the Milky Way \citep{annurev-astro-032620-021917, cooper2023overview}. Central to this pursuit is the six-dimensional phase space information, encompassing both the positions and velocities of various galactic constituents, such as stars, clusters, tidal streams, and satellites \citep{katz2018gaia, malhan2018ghostly, cantat2022milky, drimmel2023gaia}. Proper motion provides a tangential component of the three-dimensional velocity vector, which is the observed position change of a celestial object over time relative to the background sources. It is crucial for illuminating the assembly history and dynamic evolution of our Galaxy.

The measurement of proper motion involves the comparative analysis of celestial coordinates for stars captured at different epochs. Early proper motion catalogs were established primarily through the use of multiple exposures from the same survey over varying time periods. These surveys either spanned nearly the entire sky with relatively shallow observations \citep[e.g.][]{hog2000tycho,Zacharias2000,Monet2003} or targeted smaller yet more intensively scrutinized regions of the sky \citep[e.g.][]{hambly2001supercosmos,2004AJ....127.3034M}. The advent of both wide-field and highly sensitive photometric surveys subsequently enabled us to determine star proper motions more precisely using long-time baselines between different surveys \citep[e.g.][]{2021MNRAS.501.5149Q,2017ApJS..232....4T,2015PASP..127..250P}.

Measurement of proper motion necessitates the determination of a star motion relative to a well-defined velocity reference frame. Broadly speaking, there exist two fundamental strategies to align proper motions with an inertial frame, as outlined by \citep{2017ApJS..232....4T}. The first approach involves leveraging a high-precision catalog already firmly anchored to the International Celestial Reference System (ICRS) and subsequently incorporating faint sources into this frame. Such catalogs are exemplified by Tycho-2 \citep{hog2000tycho}, PPMXL \citep{2010AJ....139.2440R}, and UCAC4 \citep{2013AJ....145...44Z}. Alternatively, one may construct a reference frame by harnessing distant celestial objects whose proper motions are negligible, such as galaxies and quasars. This methodology entails meticulous cross-calibration of observations obtained at different epochs. The SDSS proper motion catalog \citep{2004AJ....127.3034M}  and the XPM catalog \citep{2009MNRAS.393..133F} were assembled following this approach. 

Employing the South Galactic Cap $U$-band Sky Survey \citep[SCUSS;][]{zou2016AJ....151...37Z}, \cite{2015PASP..127..250P} determined absolute proper motions for some 7.7 million celestial bodies via astrometric comparison with the Sloan Digital Sky Survey \citep[SDSS;][]{york2000sloan}. They implemented the Absolute Proper Motions Outside the Plane (APOP) technique \citep{2015AJ....150..137Q} to construct a reference frame utilizing background galaxies, thereby mitigating systematic errors associated with position, magnitude, and color. The SCUSS proper motion catalog records systematic errors of 0.08 mas yr$^{-1}$ and 0.06 mas yr$^{-1}$ for respective components. \cite{2017ApJS..232....4T} computed stellar proper motions using data from Gaia DR1 \citep{2016A&A...595A...2G, lindegren2016gaia}, Pan-STARRS1 \citep[PS1;][]{chambers2016pan}, SDSS, and 2MASS \citep{2006AJ....131.1163S}. They established a reference frame based on galaxy positions within PS1 and derived proper motions by examining star positions across up to nine distinct epochs in those surveys. GPS1 contains the proper motions of 350 million sources down to $m_{r}$ $\sim$ 20 and exhibits small systematic errors ($<$0.3 mas yr$^{-1}$) and high precision ($\sim$1.5 mas yr$^{-1}$). 

The trailblazing Gaia mission unveiled its third dataset in 2023 \citep{2023A&A...674A...1G}, exponentially boosting the count of celestial objects with precisely determined proper motions to an impressive 1.5 billion. Among stars brighter than 20.8 mag, a remarkable 97\% exhibit typical uncertainties in proper motion better than 1.5 mas yr$^{-1}$. Nonetheless, current Gaia proper motion measurements remain constrained by their limiting magnitudes. Wide and deep photometric surveys hold the potential to compile proper motion catalogs reaching deeper into the realm of faint stars, thereby facilitating investigations of the farthest extents and the kinematics of more distant populations in our Galaxy. The legacy imaging surveys of the Dark Energy Spectroscopic Instrument \citep[DESI;][]{2019AJ....157..168D} deliver optical imaging data with 2--3 magnitudes deeper than SDSS \citep{fukugita1996sloan, hogg2001photometricity}. Spanning nearly the entire SDSS sky coverage, these surveys present a substantial temporal baseline of approximately a decade. In this paper, we capitalize on these rich resources by merging the two photometric datasets to yield a proper motion catalog encompassing stars dimmer than the magnitude limit imposed by Gaia. 

Our catalog acts as an essential resource for augmenting the proper motions of fainter stars that lie beyond those listed in the Gaia datset. With this catalog, we can investigate hypervelocity stars, halo structures, stellar streams, and dwarf satellites. Despite the relatively high random errors associated with the DESI-SDSS proper motions, these measurements can still yield valuable statistical insights, particularly when examining groups of member stars within stellar streams and dwarf satellite galaxies. They contribute to improving the detection of these member signals alongside fainter stars and facilitate the identification of dim substructures and possibly undiscovered satellites in the halo. The catalog can be also used to gather a large sample halo stars at distances greater than 20 kpc. The abundance of potential halo stars may aid in revealing substructures and kinematic asymmetry within the halo. Additionally, we can identify fainter hypervelocity stars that lie beyond the Gaia limit. All these can enhance our understanding of the evolution history of the Milky Way. From another angle, our catalog can assist in detecting stellar streams and satellite galaxies using imaging data alone. By enabling the exclusion of foreground contamination and filtering out stars whose proper motions do not conform to the expected kinematics of halo populations, substructures, streams, and dwarf galaxies, we can refine the selection process for these objects.

The structure of this paper is outlined as follows: Section \ref{sec:2} describes the SDSS and DESI imaging data utilized in the derivation of the proper motion catalog. Section \ref{sec:3} elucidates the methodology employed for the catalog construction. Section \ref{sec:4} presents the systematic errors and precision of the derived proper motions as gauged through diverse celestial objects, encompassing galaxies, quasars, and distant halo stars, and comparisons with other established proper motion catalogs. Section \ref{sec:5} discusses the limitations inherent in our catalog and provides an application example for the determination of proper motions in star clusters situated within the catalog sky coverage. Section \ref{sec:6} gives the summary. Throughout this paper, we adopt the notation $\alpha_{*}$ and $\delta$ to represent the right ascension and declination, respectively. Furthermore, we employ $\alpha_{*}$ to signify the right ascension in the gnomonic projection coordinate system, such as $\mu_{\alpha_{*}}$ $=$ $\mu_{\alpha_{*}}\cos(\delta)$, and $\sigma_{\alpha_{*}}=\sigma_{\alpha_{*}}\cos(\delta)$. In addition, all the magnitudes are in AB mag. 

\section{Data} \label{sec:2}
\subsection{Sloan Digital Sky Survey} 
The SDSS is one of the most successful and influential projects, characterized by both photometric and spectroscopic surveys, using a 2.5-meter telescope. This telescope is instrumented by a wide-area, multiband CCD camera and multi-object fiber-fed spectrographs \citep{2006AJ....131.2332G}. The SDSS survey covers roughly one-third of the celestial sphere. It has performed concurrent observations across five photometric bands of $u$, $g$, $r$, $i$, and $z$, with a broad wavelength coverage ranging from 3000 to 11,000 \AA. The 5$\sigma$ limiting magnitudes for these bands are respectively 22.1, 23.2, 23.1, 22.5, and 20.8 mag. The saturation magnitudes are approximately 13.5, 15.0, 15.5, 15.0, and 14.0 mag in $u$, $g$, $r$, $i$, and $z$ bands, respectively \citep{1998AJ....116.3040G}. The typical astrometric uncertainties for brighter stars with $r< 19.0$ are in the range of 20 to 30 mas \citep{2002AJ....123..485S}.  

Since commencing regular operations in 2000, the SDSS has progressed through multiple stages. During SDSS-I (2000-2005), the primary objective was to conduct a "Legacy" survey encompassing five-band imaging and spectroscopy of well-targeted galaxy and QSO samples \citep{york2000sloan}. Following this, SDSS-II (2005-2008) completed the Legacy survey and further explored the three-dimensional clustering of one million galaxies and 100,000 quasars \citep{abazajian2009seventh}. Subsequently, SDSS-III, IV, and V (2008-present) have predominantly concentrated on spectroscopic observations. SDSS-III expanded the imaging coverage by adding an extra 2395 deg$^2$ in the Southern Galactic Cap (SGC), bringing the cumulative imaging area to approximately 14,500 deg$^2$ \citep{aihara2011eighth}.

In SDSS, the astrometric measurements are primarily based on the positions of objects in the $r-$band, which are obtained by fitting centroids on $r$ band \citep{pier2003astrometric}. These positions are calibrated using bright stars detected in SDSS and cross-referenced with the UCAC astrometric catalogs. The SDSS imaging runs generally overlap with each other, resulting in multiple detections of the same sources. To manage these overlaps and ensure consistent measurements, SDSS employs two key algorithms: the "window" algorithm and the "resolve" algorithm. The "window" algorithm selects the appropriate imaging run for the primary detections of objects in a given sky region. Meanwhile, the "resolve" algorithm identifies the "primary" set of detections and distinguishes them from "secondary" detections of the same sources \citep{aihara2011eighth}.

The photometric data from SDSS Data Release 12 used in this study corresponds to the ""primary" detections, which are derived from the first or best observation for each object, meaning that the astrometric coordinates (RA and DEC) are taken from a single exposure, not from a combination of multiple exposures. For sources located within the same SDSS field, their astrometric positions are consistently linked to the same epoch, as the field serves as the smallest unit where the "window" and "resolve" algorithms are implemented. Each SDSS field encompasses an area of 10\arcmin$\times$13{\arcmin} on the sky. In practice, SDSS imaging runs generally cover a continuous area of the sky under consistent observational conditions, which allows sources in nearby areas to often share the same epoch. Only at the boundaries of different SDSS imaging runs might two close-by sources have "primary" detections from exposures captured at different epochs.

\subsection{DESI Legacy Imaging Surveys}
The DESI is designed to undertake a large-scale spectroscopic redshift survey using the 4-meter Mayall telescope. It is capable of simultaneously observing 5,000 celestial objects and plans to obtain about 40 million spectra. The main goal is to explore the structure growth and expansion history of the universe \citep{2016arXiv161100036D}. 

The DESI legacy imaging surveys (hereafter LS) serve as the source for spectroscopic targets in the DESI project \citep{2019AJ....157..168D}. These surveys unite three publicly accessible optical survey programs, including the Beijing-Arizona Sky Survey (2015-2019) \citep[BASS;][]{2017PASP..129f4101Z}, the Mayall z-band Legacy Survey (2016-2018) \citep[MzLS;][]{2016AAS...22831702S}, and the Dark Energy Camera Legacy Survey (2014-2019) \citep[DECaLS;][]{2016AAS...22831701B}. In this study, we utilize the Data Release 9 released in 2021, which also incorporates imaging data from the Dark Energy Survey \citep[DES;][]{2016MNRAS.460.1270D}. The survey footprint spans a total sky coverage of approximately 20,000 square degrees. The imaging data from this release present 5$\sigma$ limiting magnitudes of 24, 23.5, and 22.9 mag for $g$, $r$, and $z$ bands, respectively.  

The astrometry of each LS exposure is tied to Gaia Data Release 2 \citep{2018A&A...616A...1G}.  Typically, for stars with $r \sim 21.0$, the astrometric uncertainties are less than 20 mas. Initially, each individual exposure undergoes an independent astrometric calibration using the Gaia DR2 reference catalog. This preliminary calibration ensures that the positional measurements are consistent and mitigates systematic errors arising from differences in the observational conditions across epochs. Specifically, the coordinates of Gaia sources are first adjusted to correspond to the epoch of the DESI exposures by employing Gaia proper motions. Afterward, the DESI positions in each exposure are aligned with the expected Gaia positions.

Sources in the LS are identified using stacked images created from weighted sums of exposures, either within individual bands or across multiple bands. These stacked images enhance the sensitivity and reliability of source detection. To obtain precise positions for the detected sources, parametric models using \textit{The Tractor}\footnote{\url{https://github.com/dstndstn/tractor}}. Unlike a straightforward weighted average of the positions from individual exposures, \textit{The Tractor} performs global fitting at the pixel level. It simultaneously models each source using all available images, convolving parametric source models (e.g., delta function, exponential, de Vaucouleurs or composite) with the specific point spread function (PSF) of each exposure. By minimizing residuals across all exposures, this process ensures that the final positions encapsulate the combined constraints from all exposures, effectively accounting for variations in observational conditions, such as seeing \citep{2019AJ....157..168D}. Consequently, The coordinates in the DESI LS are derived through a comprehensive process that leverages data from multiple exposures to achieve high precision.

This methodology leverages the strengths of multi-exposure data, reducing the impact of noise or systematic offsets that might be present in individual images. Since the final coordinates are determined from a multi-epoch combined model, it is necessary to adopt an average epoch for these coordinates. While \textit{The Tractor} does not explicitly provide the mean observation epoch for the fitted sources in its output catalog, it does report the range of observation epochs for each source. To estimate the mean epoch, we use the average exposure time of all images utilized in \textit{The Tractor} processing, as provided in the additional observation file\footnote{\url{https://www.legacysurvey.org/dr9/files/}}. However, since the final positions are not obtained through a simple averaging of individual exposure positions, using the average observation time to represent the mean epoch is also not entirely accurate. The position fitting process further complicates the calculation of the mean epoch. Although this estimation introduces some uncertainty, the long time baseline between DESI and SDSS observations significantly mitigates its potential impact on the results. Therefore, we use the average of all observation times as a first-order approximation of the weighted mean epoch. Future studies on proper motions based on individual DESI exposures will resolve this issue.

\begin{figure}[htbp]
    \centering
    \includegraphics[width=0.48\textwidth]{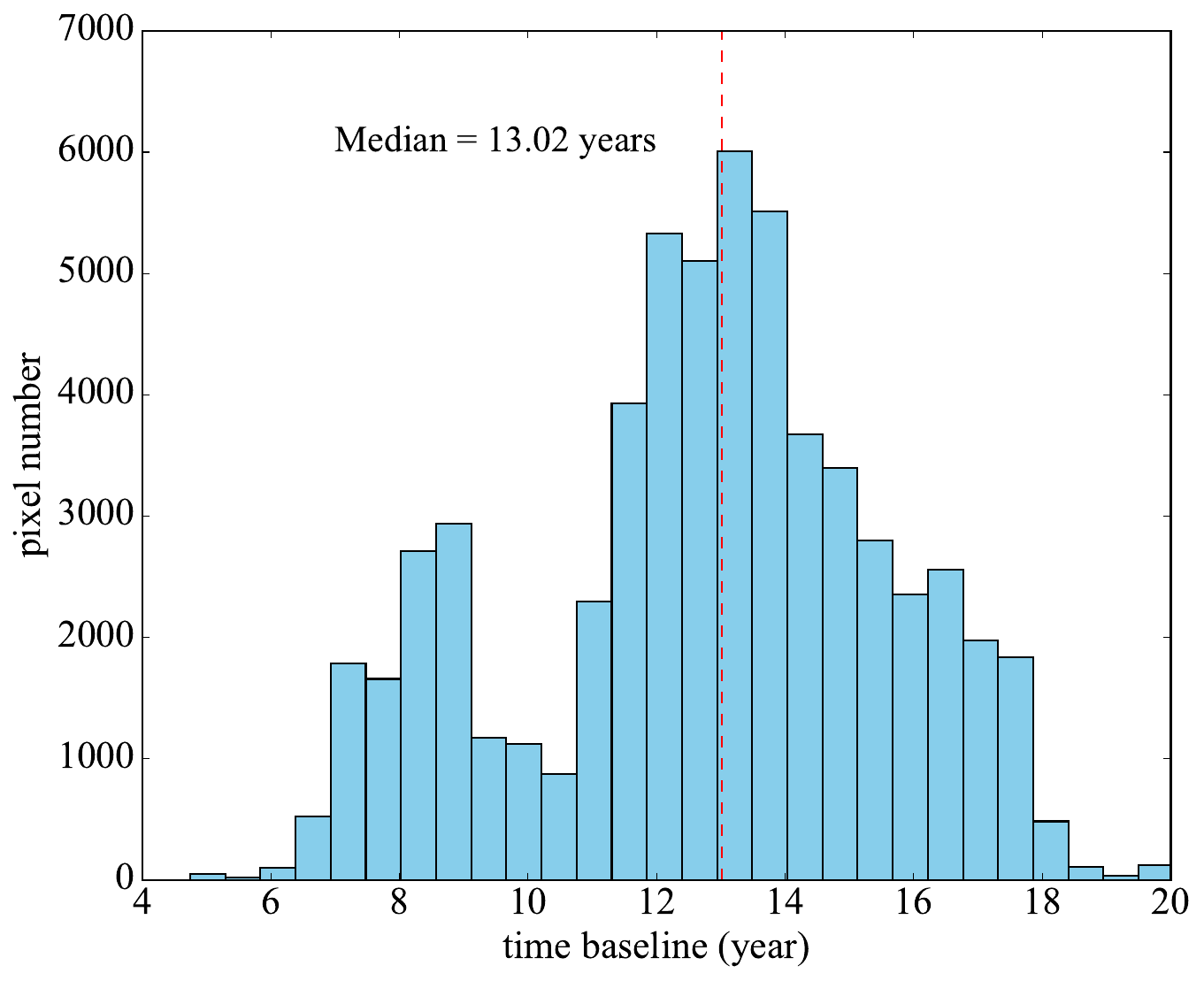}
    \caption{Distribution of the time intervals for celestial sources between the SDSS and DESI observations. The red dashed line represents the median value.}
    \label{fig:time}
\end{figure}

\begin{figure}[htbp]
    \centering
    \includegraphics[width=0.48\textwidth]{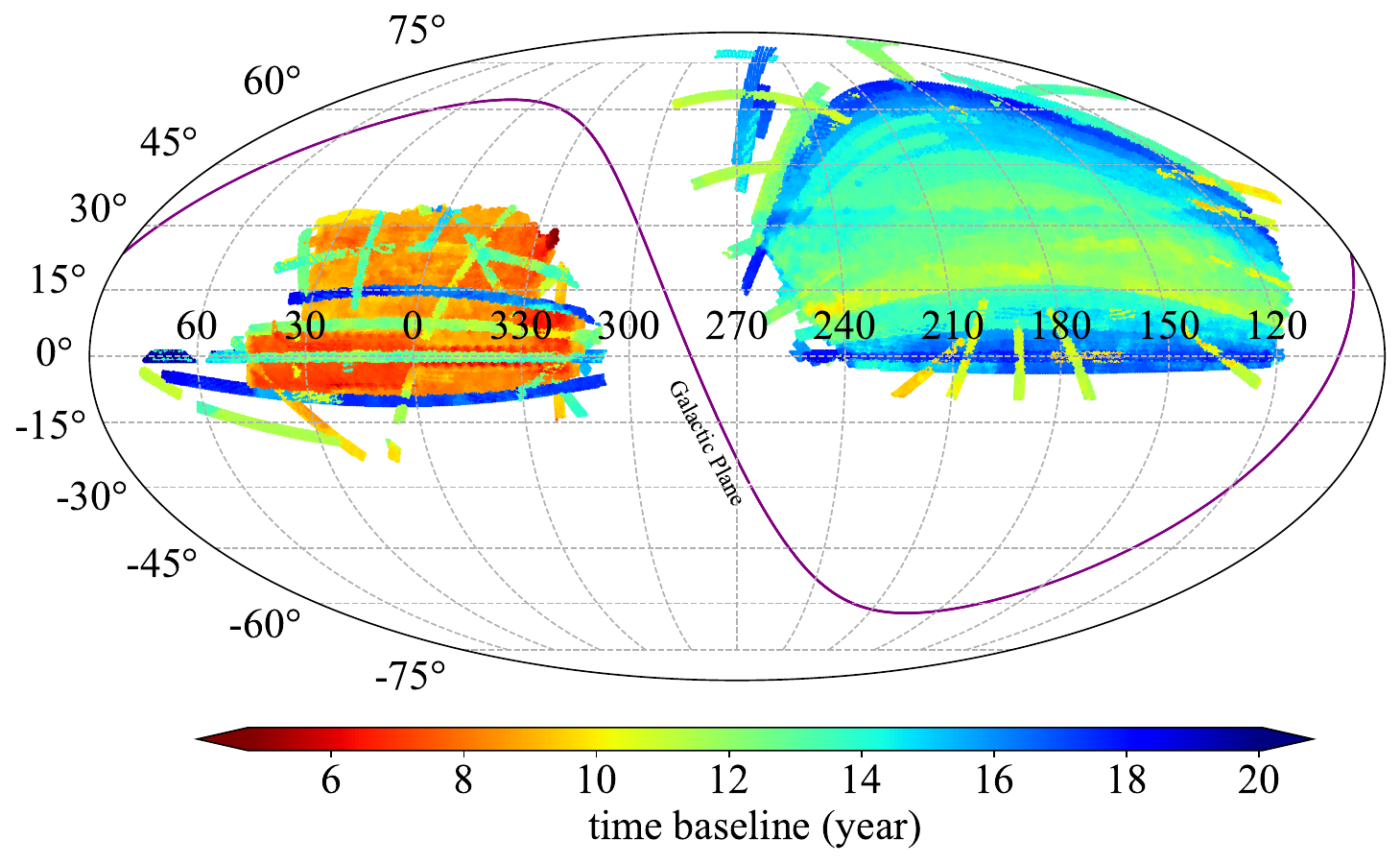}
    \caption{Spatial distribution of the time interval between SDSS and DESI observations. The purple curve represents the Galactic plane.}
    \label{fig:time_distribution}
\end{figure}

The temporal difference between SDSS and DESI observation epochs is calculated. Figure \ref{fig:time} shows the histogram of the time interval. This time interval varies broadly from a minimum of 5 years to a maximum of 19 years, with a median value of about 13 years. Such a lengthy time baseline significantly contributes to reducing inherent errors in proper motion estimation. There are two peaks in the time interval distribution, which correspond to early observations by SDSS-I \& II and more recent observations in the SGC by SDSS-III, respectively. This difference in time intervals caused by SDSS observations can be clearly seen in Figure \ref{fig:time_distribution}. 

\section{Constructing the DESI-SDSS proper motion catalog}
\label{sec:3}
Ideally, point sources are advantageous for establishing a reference frame due to their narrow Full Width at Half Maximum (FWHM), which enhances the positional measurement precision \citep{fritz2010mn}. Nonetheless, the dearth of extragalactic point sources poses a challenge for constructing such a framework. Quasars, despite being candidates, are too sparsely distributed. Alternatively, Milky Way stars with accurately measured velocities could serve this purpose, such as those stars in the \textit{Hipparcos} catalog \citep{hog2000tycho}. Unfortunately, these stars are too bright so that most of them are saturated in both SDSS and DESI observations, rendering them unsuitable for our purposes. Consequently, this work adopts galaxies as the foundation for the reference frame. The abundance of galaxies can considerably mitigate systematic errors, enhancing the accuracy of our analyses. 

For DESI, the astrometric framework directly incorporates positions from sources observed by Gaia. For sources not observed by Gaia—which are the primary focus of this study—the DESI astrometric calibration process aligns their positions with the Gaia frame through rigorous corrections for individual exposures. This calibration ensures that DESI sources are effectively integrated into a Gaia-consistent framework, enabling the reliable use of DESI galaxies as reference objects for subsequent transformations. Although there are potential residual systematics in DESI galaxy positions, the high density of DESI galaxies can significantly reduce small-scale systematics. Future studies will further quantify these systematics using DESI single-epoch observational data.

The procedure is outlined as follows: First, we cross-match the SDSS and DESI catalogs utilizing a 1{\arcsec} matching radius, yielding a shared sky coverage of 12,589 deg$^2$, which contains approximately 260 million objects. Taking into account the unique DESI photometric strategy that employs Gaia coordinates directly for photometry, we exclude these Gaia-detected objects (around 52.6 million) from subsequent analysis. Subsequently, position-, magnitude-, and color-dependent corrections are made to eliminate any systematic differences between the two surveys. Ultimately, the corrected coordinate displacements between DESI and SDSS observations yield the final proper motion calculations.

\subsection{Correction for Position-dependent Astrometric Offsets}
\label{subsec:3.1}
Prior to proper motion computation, the selection of background galaxy samples is imperative, alongside establishing a reference frame anchored on DESI-derived coordinates for these chosen galaxies. Following this, corrections are applied to address the systematic positional discrepancies of galaxies present in the SDSS catalog. It aims to align them with the established reference frame. These positional offsets, influenced by atmospheric and instrumentation conditions, exhibit regional variability across the sky. 

In the SDSS catalog, the primary epoch is consistent within each 10\arcmin$\times$13{\arcmin} field, with variations between adjacent fields/runs typically spanning a few days. However, these variations become uncertain over larger areas, ranging from months to years depending on the observation schedule. A similar pattern is observed in the DESI catalog, where the average epoch within the DESI field of view (1.1 or 3.2 square degrees) is nearly uniform, with variations generally limited to days. In larger regions, differences can also widen significantly, potentially spanning months or years. To account for the spatially varying offset, the merged DESI-SDSS dataset is segmented into equal-area pixels using HEALPix \citep{gorski2005healpix}. HEALPix generates the non-overlapping sky divisions of equal areas, referred to as pixels, which have strict boundaries. In our analysis, the pixel serves as the smallest unit for reference-frame systematic corrections, with each object projected onto a single pixel. Each pixel has a sky area of 0.21 deg$^2$ (nside$=$128). While \citet{2017ApJS..232....4T} reported minimal variation in positional offsets and root mean square values when transitioning between 1{\arcdeg}$\times$1{\arcdeg} and 0.5{\arcdeg}$\times$0.5{\arcdeg} pixel sizes. We also examined smaller pixel sizes and found that the current configuration of nside$=$128 yields the best results. Shrinking pixel dimensions further undermine the statistical validity due to insufficient galaxy numbers per pixel. Thus, our pixel division represents an optimal balance between minimizing systematic effects and maintaining adequate statistical robustness. Although the mean epoch within most pixels is uniform when we choose nside$=$128, a small fraction of pixels ($\sim1\%$) span multiple observation regions, resulting in the presence of two or more distinct epoch distributions within the same pixel. This could potentially impact subsequent positional offset corrections. We have specifically marked these pixels to indicate this potential issue.

\begin{figure*}[htbp]
  \begin{subfigure}[t]{.48\textwidth}
    \centering
    \includegraphics[width=\linewidth]{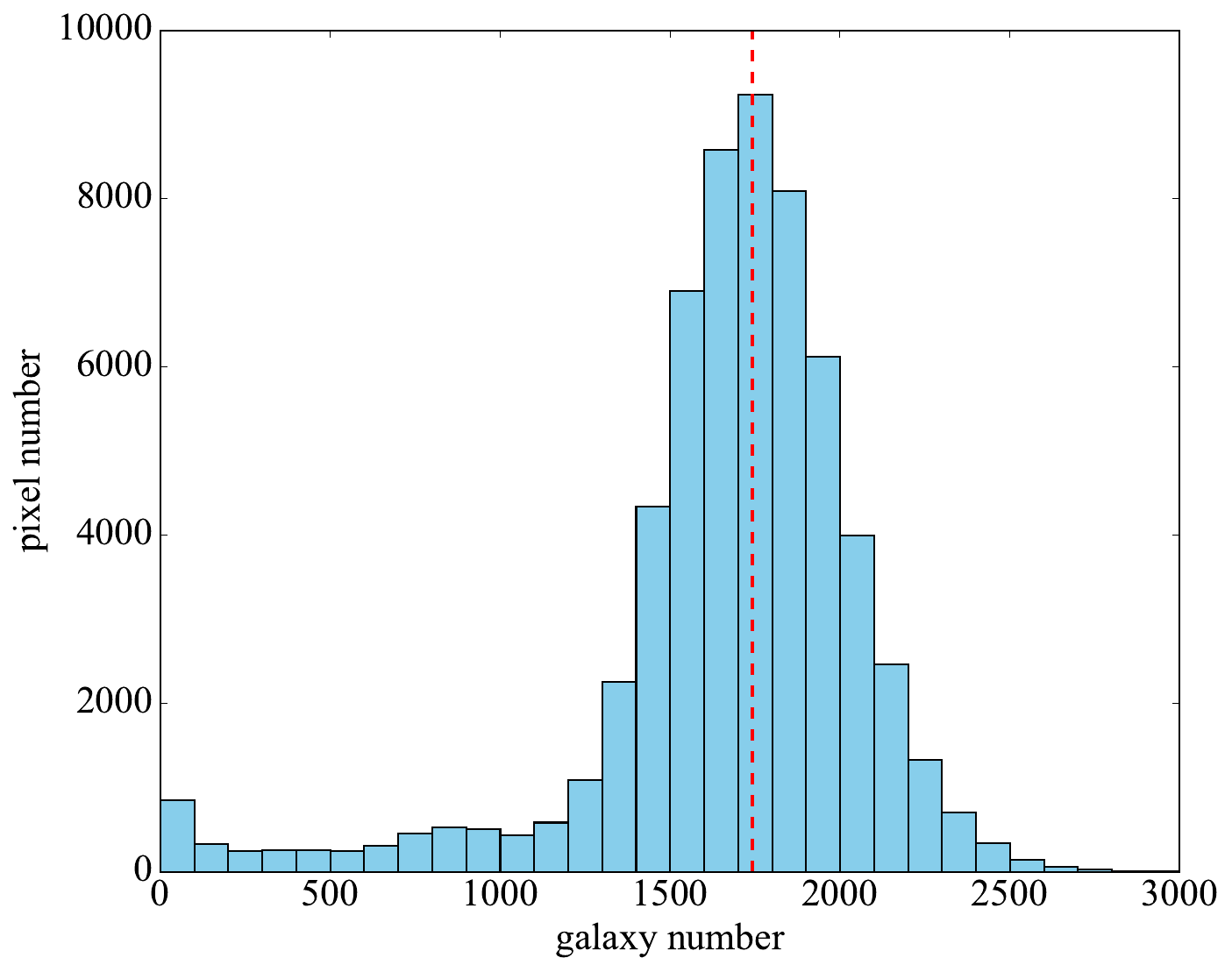}
  \end{subfigure}
  \hfill
  \begin{subfigure}[t]{.48\textwidth}
    \centering
    \includegraphics[width=\linewidth]{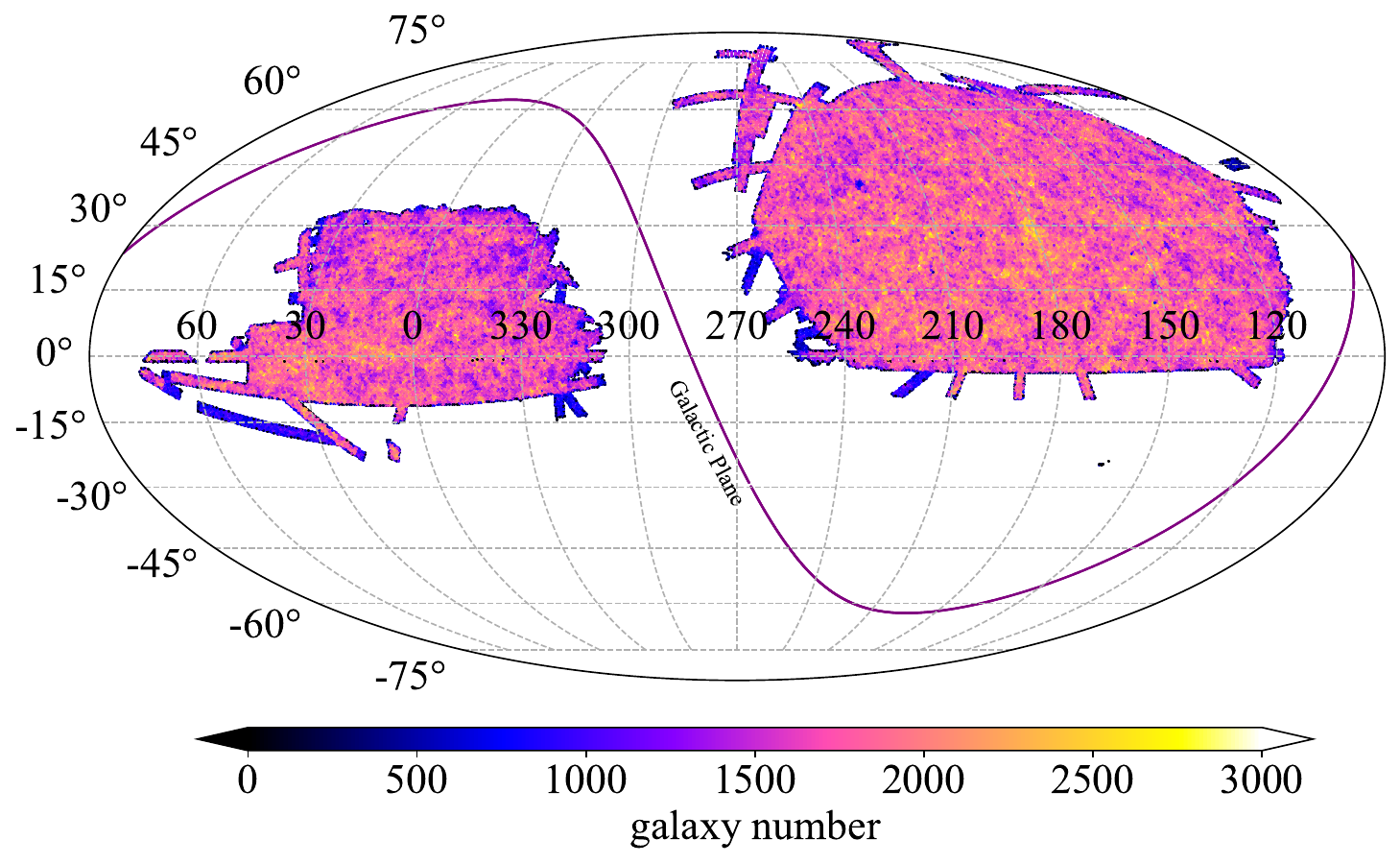}
  \end{subfigure}
 \caption{Left: histogram of galaxy numbers for all pixels. The dashed line shows the median value. Right: spatial distribution of galaxy numbers. The purple curve represents the Galactic plane.} \label{fig:all_pixel_numbers}
\end{figure*}

In the process of identifying the reference galaxy sample, we impose the following criteria for selection:
\begin{enumerate}
    \item[(1)] 16 $<$ MAG\_R $<$ 21.5 (magnitude cut),
    \item[(2)] MAGERR\_R $<$ 0.15 (magnitude error cut),
    \item[(3)] TYPE\_SDSS $= 3$ \& TYPE\_DESI $\neq$ PSF (morphological classification as galaxy in both surveys);
    \item[(4)] GAIA\_PHOT\_G\_MEAN\_MAG $=$ 0 (non-Gaia objects),
\end{enumerate}
where MAG\_R is the $r$-band magnitude measured by DESI,  MAGERR\_R is its associated error, TYPE\_SDSS and TYPE\_DESI are denoted classification parameters provided respectively by SDSS and DESI, and GAIA\_PHOT\_G\_MEAN\_MAG is the Gaia $G$-band magnitude. Applying these filters enables the isolation of non-Gaia galaxies exhibiting relatively small positional measurement uncertainties from DESI. The left panel of Figure~\ref{fig:all_pixel_numbers} shows the histogram of the galaxy numbers for all pixels. The median value is around 1750, a density that effectively curtails the position error of the reference frame to under 4 mas, prior to accounting for the time baseline.

The right panel of Figure~\ref{fig:all_pixel_numbers} presents the spatial distribution of the galaxy numbers for all pixels. In pixels hosting over 800 galaxy samples, we compute the 3$\sigma$ medians of the positional discrepancies ($\Delta \alpha_{*}$ and $\Delta \delta$) between SDSS and DESI coordinates. Nevertheless, pixels situated along the periphery of the survey footprint often contain fewer than 800 galaxies. To uphold the high precision of the reference frame, we implement a tailored approach for these marginal pixels. We integrate all the neighboring pixels surrounding the target pixel (within a 2{\arcdeg} radius) and gradually increase the search radius from 0.5{\arcdeg} with a step size of 0.1{\arcdeg} to select galaxy samples, until the number of galaxy samples reaches 800 or the radius exceeds 2{\arcdeg}.

\begin{figure*}[htbp]
    \centering
    \includegraphics[width=\linewidth]{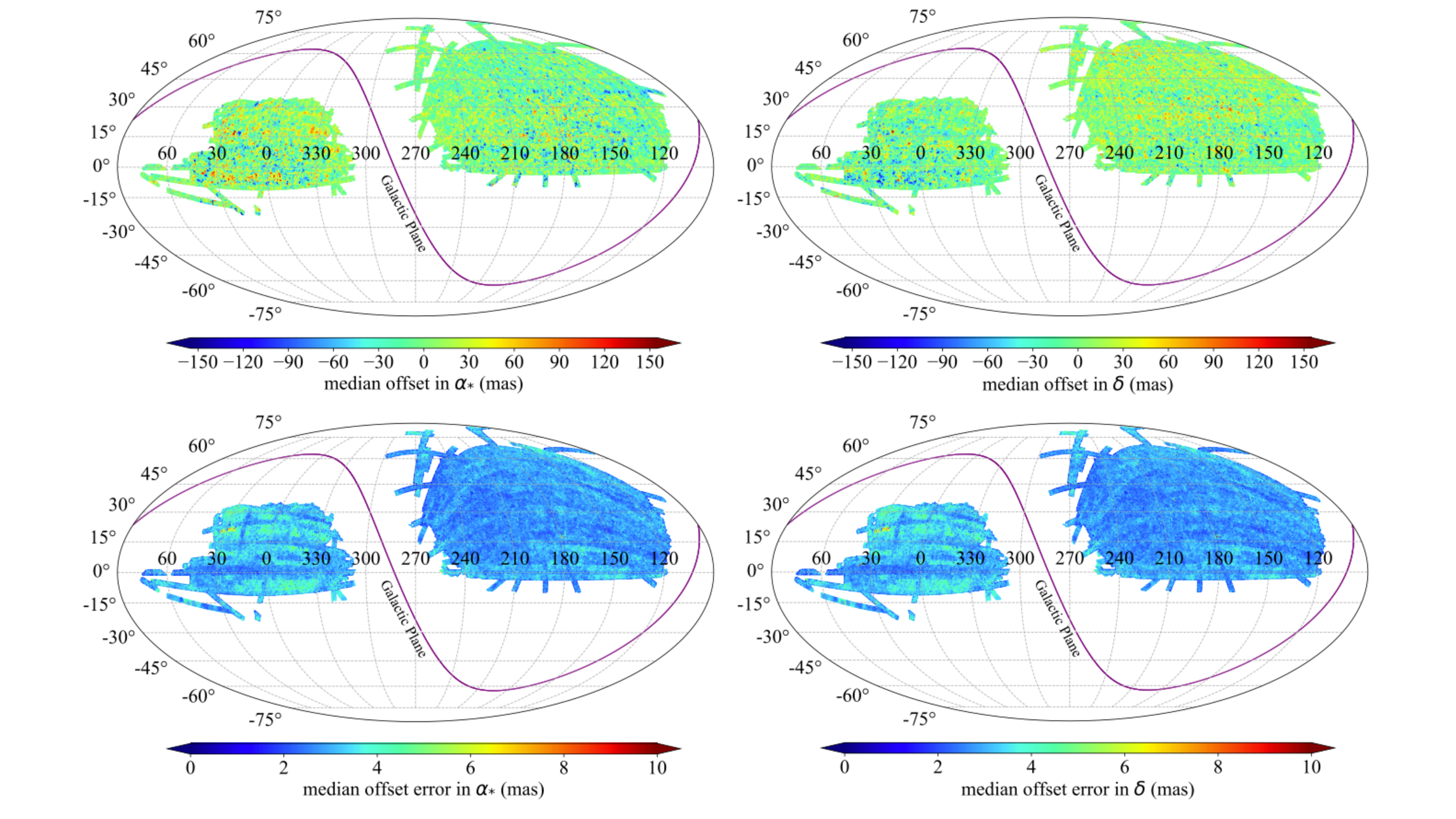}
    \caption{Spatial distribution of positional offsets (upper panels) and offset errors (lower panels). The left and right panels represent the distributions for $\alpha_{*}$ and $\delta$, respectively. The purple curves represent the Galactic plane.}
    \label{fig:position_galaxy_distribution}
\end{figure*}

The upper panels of Figure~\ref{fig:position_galaxy_distribution} illustrate the median positional offsets between the SDSS and DESI observations, while the lower panels present the error of the median offset, which is calculated as the standard deviation of the offsets divided by the square root of the galaxy number. A notable characteristic is that the majority of pixels exhibit an offset below 5 mas. Nonetheless, minor-scale fluctuations, on the order of $\pm$100 mas, are visible in the median offset patterns. We apply corrections to the SDSS source coordinates by adjusting them according to the computed median offsets in both $\alpha_{*}$ and $\delta$. 

\subsection{Corrections for color- and magnitude-dependent astrometric offsets} \label{subsec:3.2} 
Despite the positional corrections previously detailed, a residual discrepancy persists between the corrected SDSS source positions and the established reference frame. This lingering discrepancy exhibits a dependency on both the magnitude and color of sources, necessitating additional correction measurements. Its origin lies in multifaceted factors, such as atmospheric refraction, imprecise measurements for faint objects, and the differential chromatic refraction (DCR) effect \citep{2009AJ....138...19K}. Atmospheric refraction, a result of the Earth's atmosphere bending incoming light, displaces celestial sources towards the zenith. This displacement angle is influenced by atmospheric conditions and is dependent on the wavelength. The DCR-induced blurring of the point spread function (PSF) in imaging observations affects the position measurement accuracy. Given that both SDSS and DESI surveys are conducted by ground-based telescopes, the astrometry is susceptible to the DCR effect. To correct for those magnitude- and color-dependent astrometric offsets, we draw more galaxies from a broader area.

\begin{figure*}[htbp]
  \begin{subfigure}[t]{.48\textwidth}
    \centering
    \includegraphics[width=\linewidth]{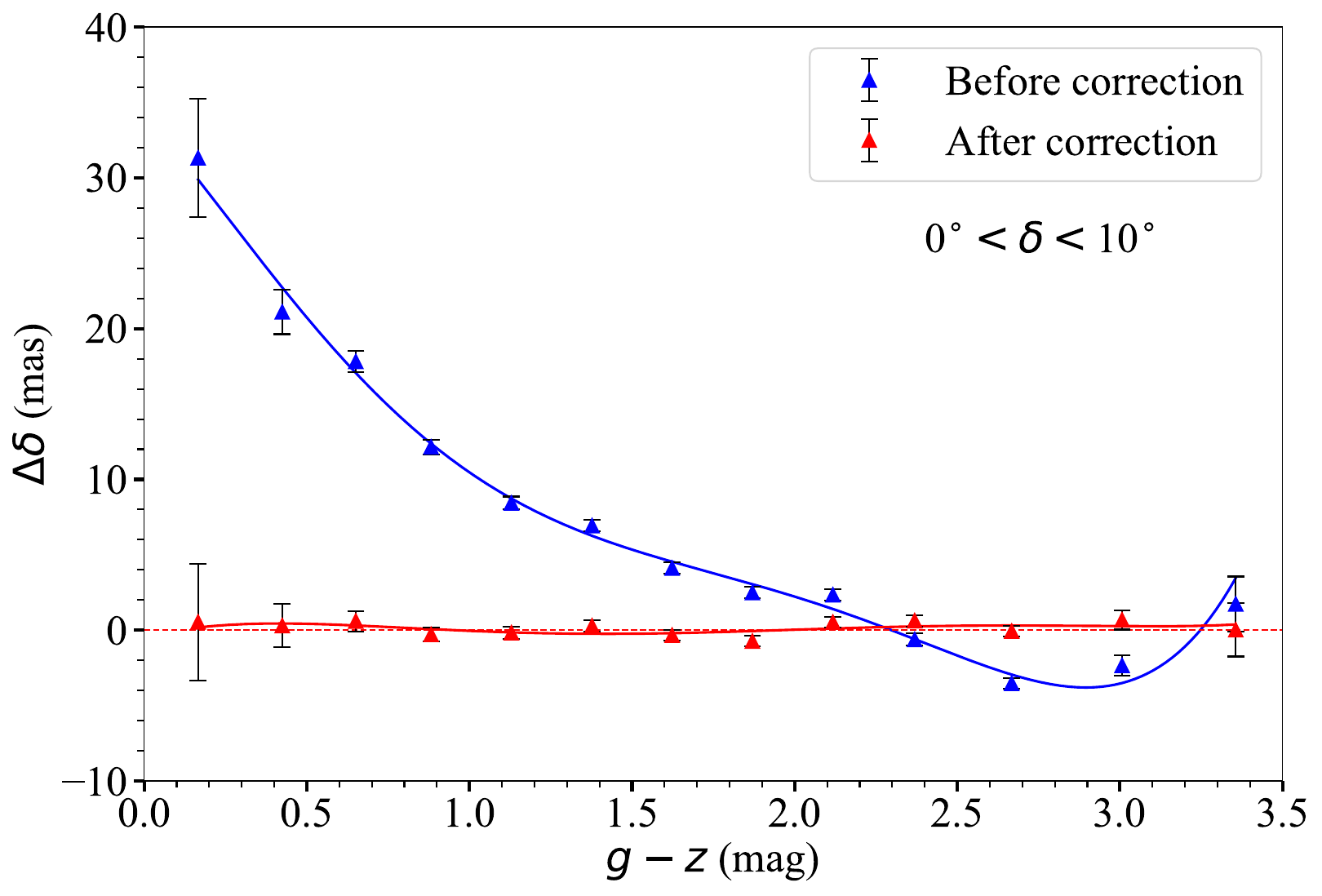}
  \end{subfigure}
  \hfill
  \begin{subfigure}[t]{.48\textwidth}
    \centering
    \includegraphics[width=\linewidth]{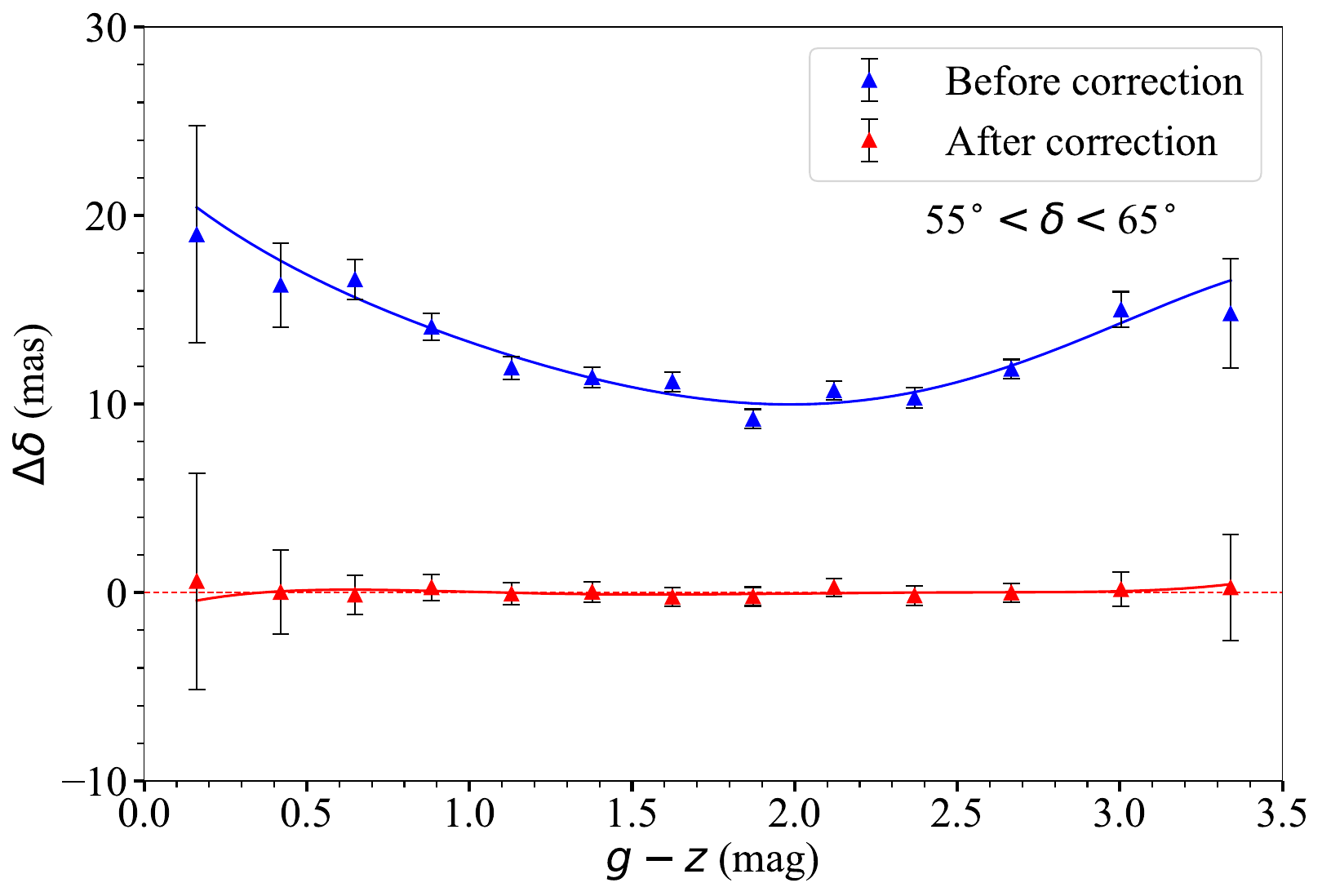}
  \end{subfigure}

  \medskip

  \begin{subfigure}[t]{.48\textwidth}
    \centering
    \includegraphics[width=\linewidth]{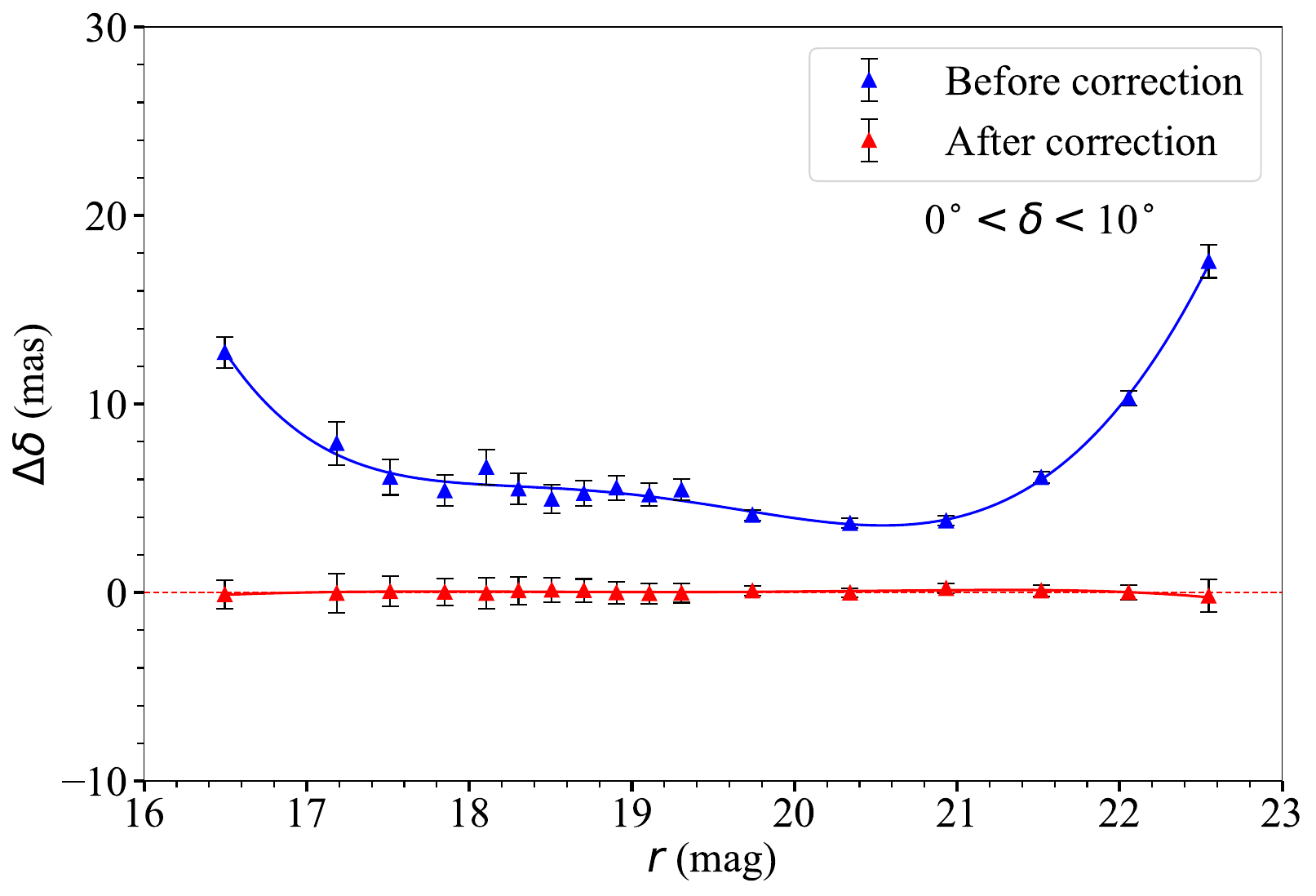}
  \end{subfigure}
  \hfill
  \begin{subfigure}[t]{.48\textwidth}
    \centering
    \includegraphics[width=\linewidth]{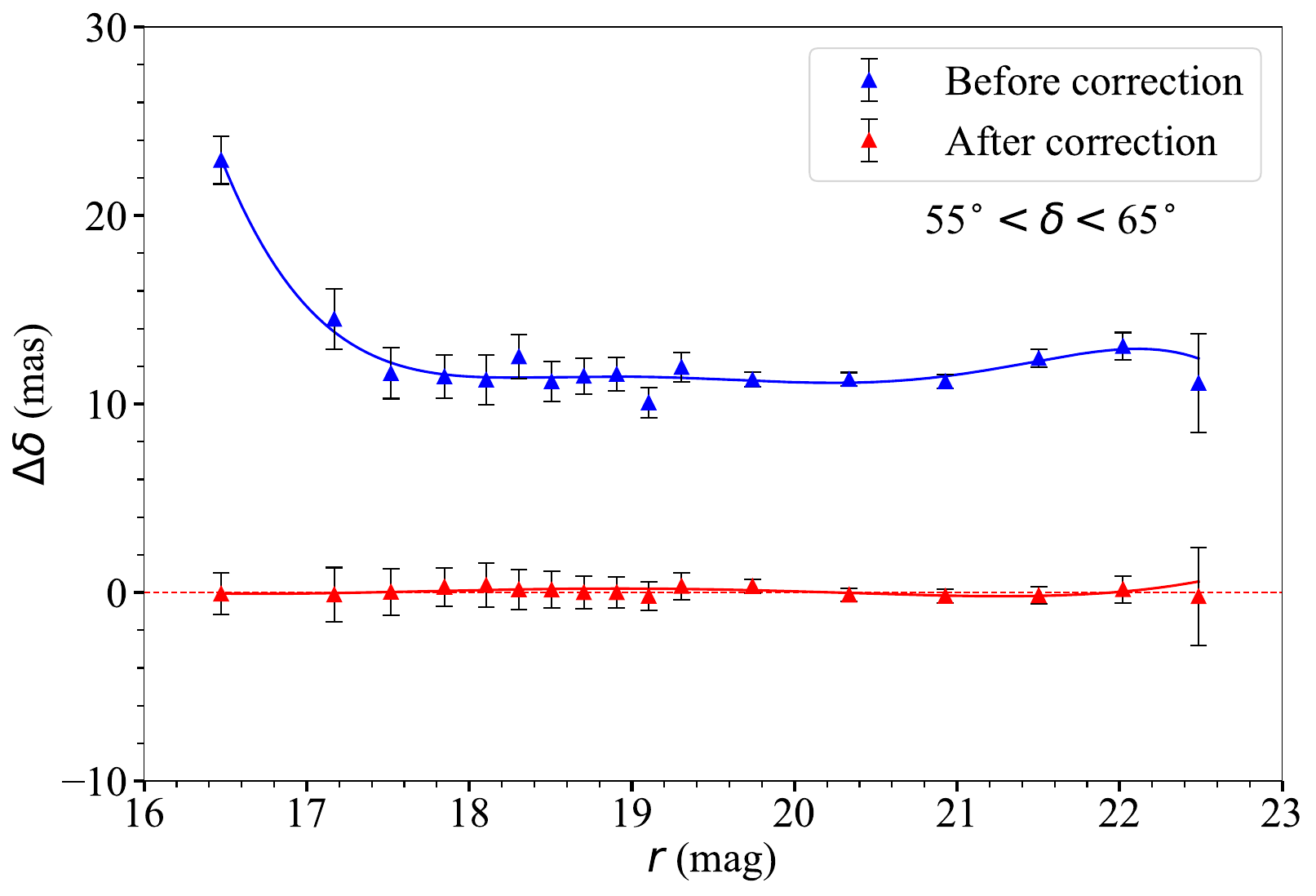}
  \end{subfigure}
  \caption{Positional offsets in $\delta$ vary with color and magnitude at different declinations. The top and bottom panels present the offsets as functions of color and magnitude, respectively. The triangles denote 3$\sigma$ median positional discrepancies observed between the SDSS recorded galaxy positions and the established reference frame in different color/magnitude bins. The error bars are the corresponding standard deviation after 3$\sigma$-clipping, reflecting the dispersion of the data points around the median. The curves are the best-fit polynomial. Blue and red represent the results before and after correction, respectively.}
  \label{fig:polyfit}
\end{figure*}

We commence by addressing the positional offset attributable to color variations. In each pixel, a reference galaxy sample is collated from contiguous pixels within a radius of 2.5$^\circ$ to 5$^\circ$. Here, we adopt the wider radius when the galaxy count within 2.5{\arcdeg} falls short of 2,000, gradually increasing the radius with a step size of 0.1{\arcdeg}. The selection criteria for the reference galaxy samples are similar to those in the process of the position-dependent correction, except enforcing a stricter magnitude error threshold (MAGERR\_R $<$ 0.1) to enhance the accuracy of color determinations. Galaxies are then segregated into multiple color bins. The upper panels of Figure \ref{fig:polyfit} show the positional offset in $\delta$ as functions of color and magnitude at different declinations. For color bins central to the figure (where 1.0 $<g-z<$ 2.5) of this figure, given the ample quantity of galaxies, they have been evenly distributed with an interval of 0.25. Conversely, at both magnitude ends, the bin sizes increment to ensure a minimum count of 200 galaxies per bin. Inside these bins, we compute the 3$\sigma$ median ($\mu$) positional offset and the standard error ($\sigma$). Further, a fifth-degree polynomial fit, inversely weighted by $\sigma$, is applied to $\mu$ as a function of the color $g-z$. From Figure \ref{fig:polyfit}, we can see that the color-related corrections exhibit a clear variation across different declinations. This confirms that we need to make color corrections in different sky positions. The resulting fits for the offsets in $\alpha_{*}$ and $\delta$ are used to correct the SDSS source coordinates. For sources that fall outside the color or magnitude boundaries, we avoid extrapolation by directly applying the median values from the boundary bins for corrections.

The correction procedure for magnitude-induced offsets is similar to that of the above color-induced correction. But we impose a different magnitude limit of 15 $<$ MAG\_R $<$ 23. The lower panels of Figure \ref{fig:polyfit} demonstrate the magnitude-dependent corrections. Except for the brightest and faintest ends, the overall correction is less than $\sim$1 mas. Through all of the aforementioned corrections, the coordinates of all SDSS sources have been realigned to coincide with the DESI reference frame. The  ultimate proper motion is calculated as the displacement between the DESI coordinate and corrected SDSS position, divided by the total span of time elapsed between the observations ($\Delta t$):
\begin{equation}
    \mu_{\alpha_{*},\delta} = \frac{\alpha_{*}, \delta_{\text{DESI}} - \alpha_{*}, \delta_{\text{SDSS}}}{\Delta t}. 
    \label{eql:1}
\end{equation}

We should notice that there might be the potential influence of galaxy morphology and structure on centroid measurements across different filters. Galaxies exhibit a wide range of morphologies and structural features, which can lead to slight shifts in centroid positions between filters due to variations in light distribution and wavelength-dependent effects. In our analysis, the impact of these filter-dependent centroid shifts is mitigated by the large sample size of galaxies used in each color correction step. Each calibration process involves thousands to tens of thousands of galaxies, allowing random centroid shifts due to morphological or structural variations to average out statistically. This significantly reduces the overall effect of filter-dependent centroid shifts as a systematic bias.

\section{The precision analyses}
\label{sec:4}
Various celestial objects, including galaxies, quasars, remote stars, and sources in other proper motion catalogs can be used to verify the precision of our derived proper motions. Galaxies and quasars, due to their immense distances, exhibit near-zero proper motion. Thus, the mean and standard deviation of their measured proper motions can directly quantify the systematic and random errors inherent in our measurements. Nevertheless, galaxies, being extended and potentially asymmetric entities, suffer from reduced positional measurement accuracy. Moreover, quasars with pronounced emission lines can present unique challenges related to DCR. Halo stars, though distant, still exhibit measurable proper motion that cannot be disregarded. Consequently, each validation methodology boasts its own merits and constraints, warranting individual examination and discussion.

\subsection{Galaxies}
\label{sec:4.1}
\begin{figure*}[htbp]
  \begin{subfigure}[t]{.48\textwidth}
    \centering
    \includegraphics[width=\linewidth]{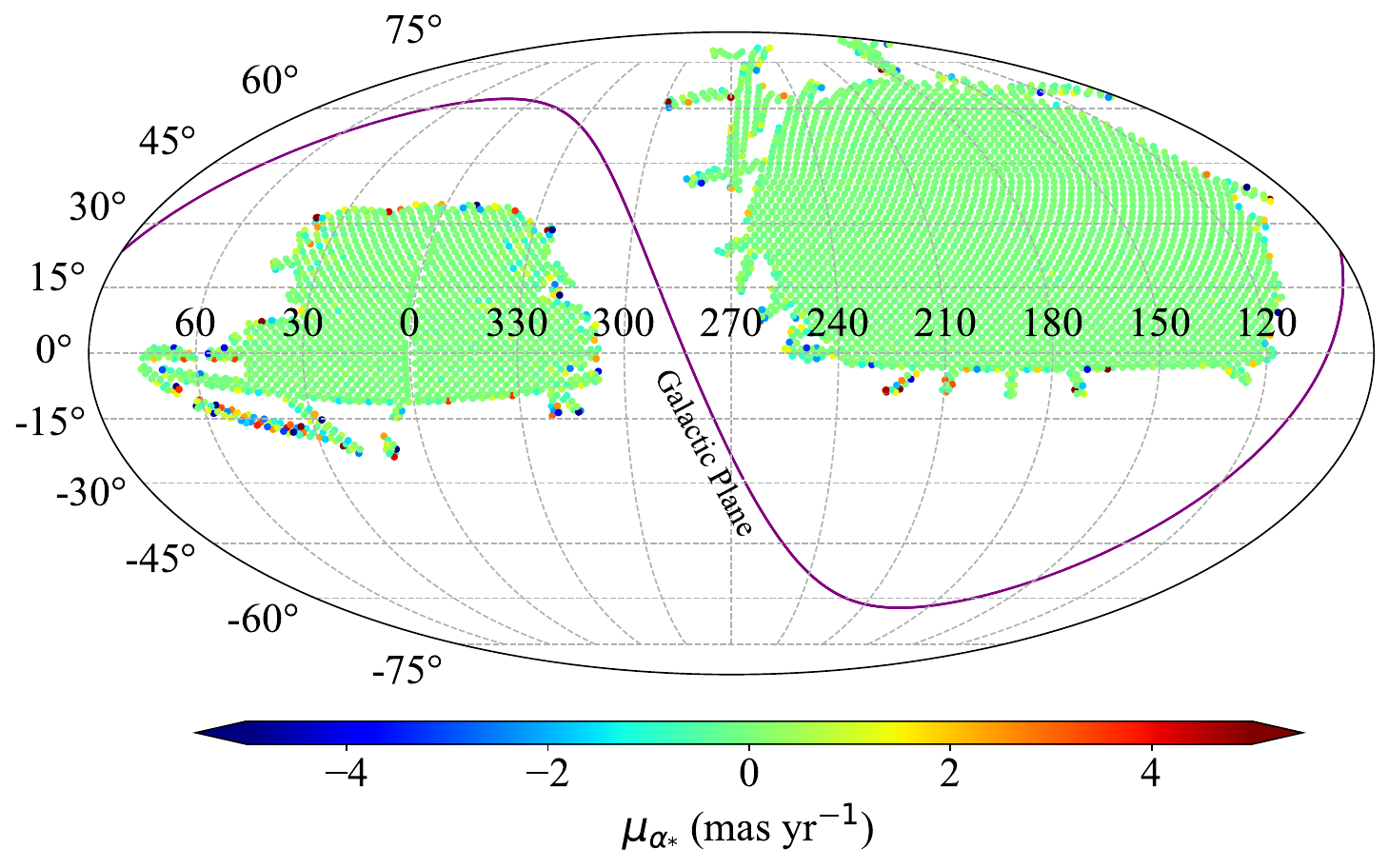}
  \end{subfigure}
  \hfill
  \begin{subfigure}[t]{.48\textwidth}
    \centering
    \includegraphics[width=\linewidth]{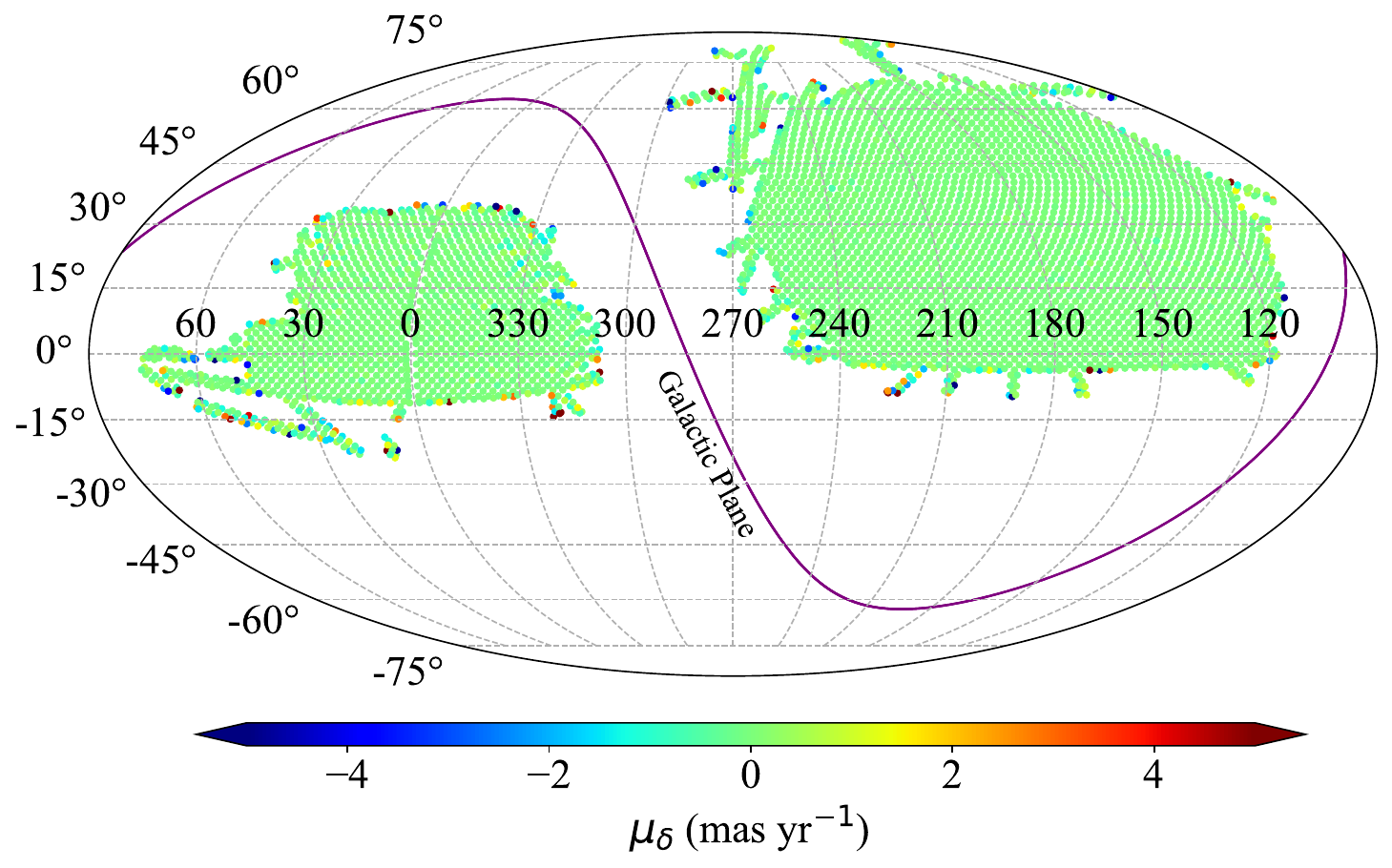}
  \end{subfigure}

  \medskip

  \begin{subfigure}[t]{.48\textwidth}
    \centering
    \includegraphics[width=\linewidth]{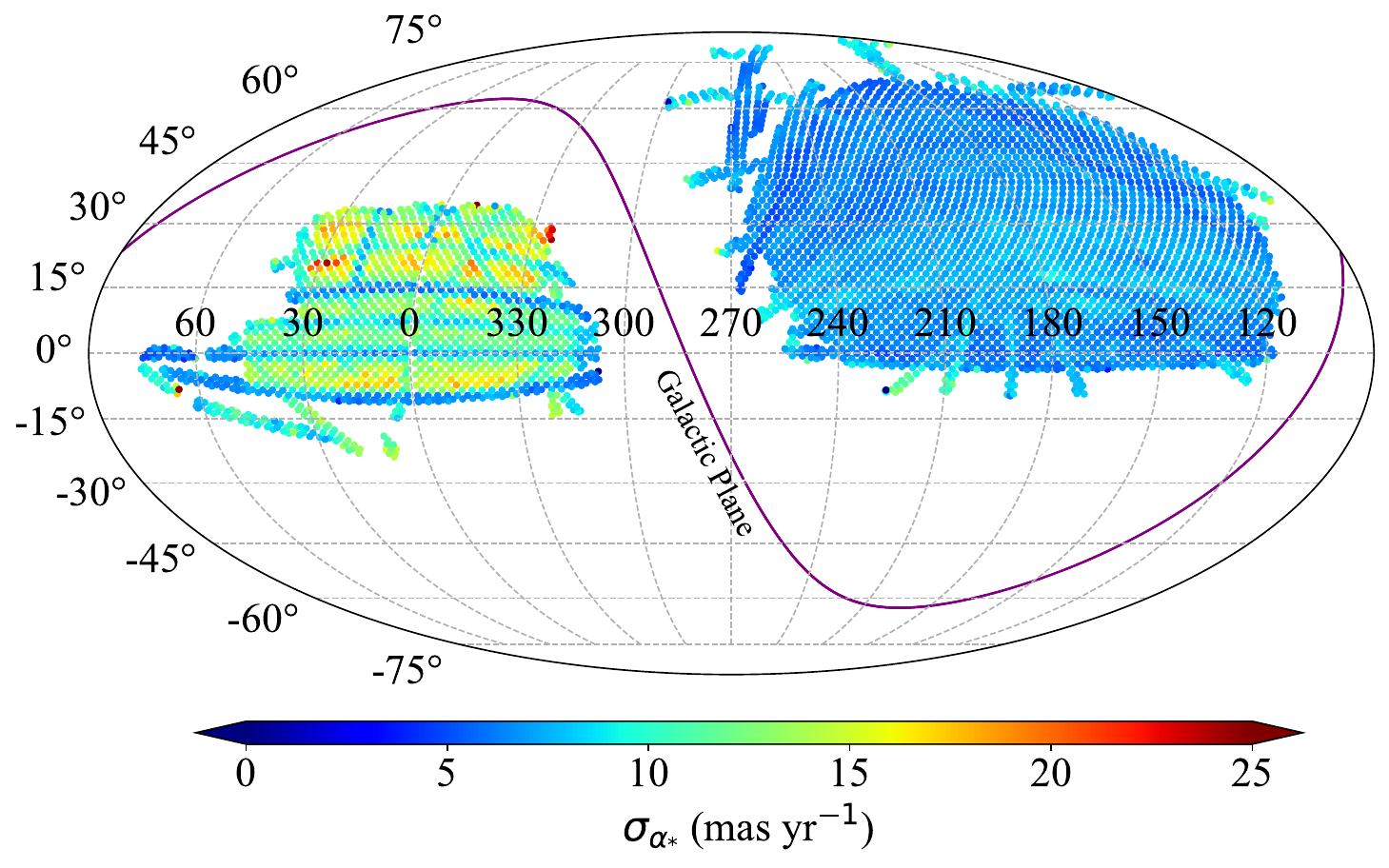}
  \end{subfigure}
  \hfill
  \begin{subfigure}[t]{.48\textwidth}
    \centering
    \includegraphics[width=\linewidth]{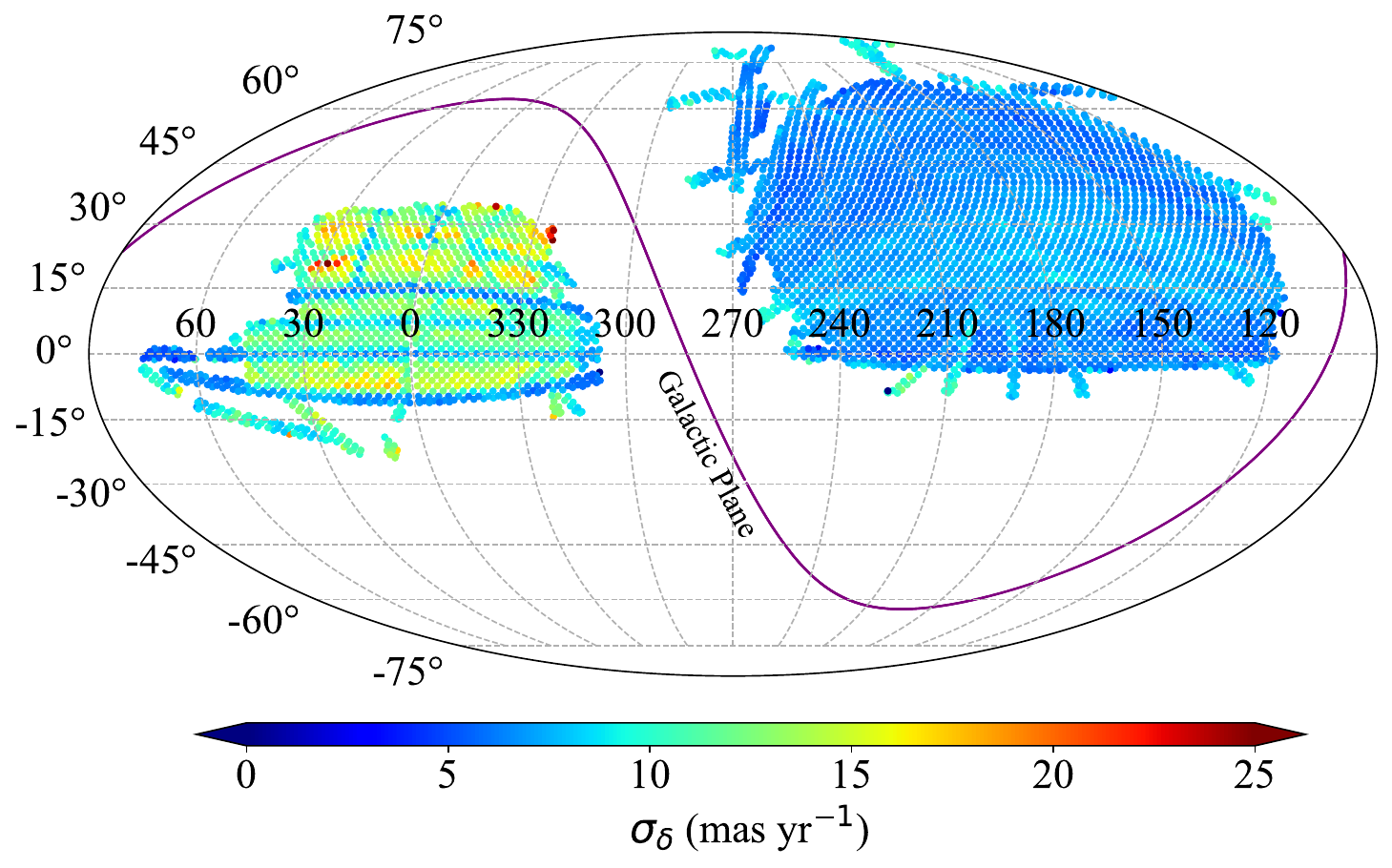}
  \end{subfigure}
  
  \caption{Spatial distribution of the systematic (top panels) and random errors (bottom panels) with respect to galaxies. The left and right panels show the errors for $\alpha_{*}$ and $\delta$, respectively. The purple curves represent the Galactic plane.} \label{fig:galaxy_distribution}
\end{figure*}

The measured proper motions of galaxies with 16 $<r<$ 21.5 can be used to evaluate the systematic and random errors. Figure \ref{fig:galaxy_distribution} shows the spatial distribution of the systematic and random errors. As depicted in this figure, the vast majority of systematic errors are confined beneath 1 mas yr$^{-1}$. It is noteworthy that galaxies, being extended objects, are prone to having larger positional uncertainties than point sources, consequently inflating the random errors associated with their inferred proper motions. Moreover, the data reveal a pattern in SGC with discernibly heightened random errors. This pattern perfectly aligns with the sequential observation scheme of the SDSS observing runs, as illustrated in Figure \ref{fig:time_distribution}. We also compare the SDSS and Gaia data and confirm its existence. The shorter time span between observations is detrimental to the accuracy of proper motion measurements. 

\subsection{Quasars}
\label{sec:4.2}
Consistent with the galaxy analysis, we evaluate both systematic and random errors in proper motion utilizing a dataset comprising 734,012 spectroscopically verified quasars from the SDSS. We also find that the random errors in the SGC are significantly higher than those ($\sim$ 3 mas yr$^{-1}$) in the NGC. Given the high positional accuracy achievable with point-like quasars, the resultant random errors in their inferred proper motions are considerably diminished relative to those of galaxies.

\begin{figure}[htbp]
    \centering
    \includegraphics[width=0.42\textwidth]{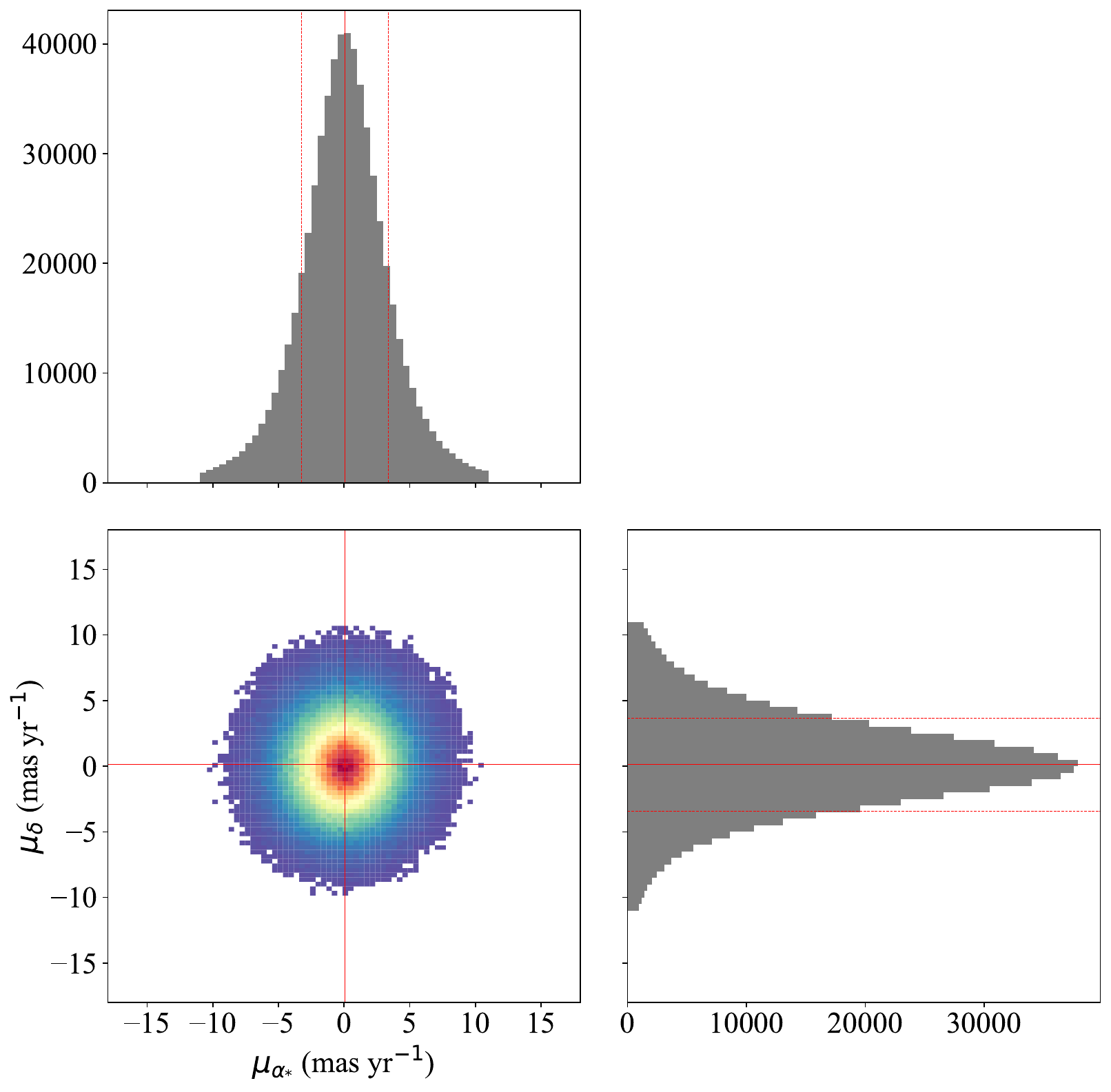}
    \caption{Distribution of proper motions for quasars. The top and right subplots present the histograms of $\mu_{\alpha_{*}}$ and $\mu_\delta$, respectively. The red lines denote the median and standard deviation values for both the $\alpha_{*}$ and $\delta$ components.}
    \label{fig:quasar_distribution}
\end{figure}

The proper motion characteristics of quasars with 16 $<r<$ 21.5 are in good agreement with Gaussian distributions, as illustrated in Figure~\ref{fig:quasar_distribution}. Consequently, they serve as a reliable basis for estimating the systematic and random errors present in our DESI-SDSS proper motion catalog. The median values of the proper motion components $\mu_{\alpha_{*}}$ and $\mu_{\delta}$ are estimated to be 0.06 mas yr$^{-1}$ and 0.12 mas yr$^{-1}$, respectively. The standard deviation around these median values (random error) is measured at 3.29 mas yr $^{-1}$ for $\alpha_{*}$ and 3.55 mas yr$^{-1}$ for $\delta$.

\begin{figure*}[htbp]
  \begin{subfigure}[t]{.48\textwidth}
    \centering
    \includegraphics[width=\linewidth]{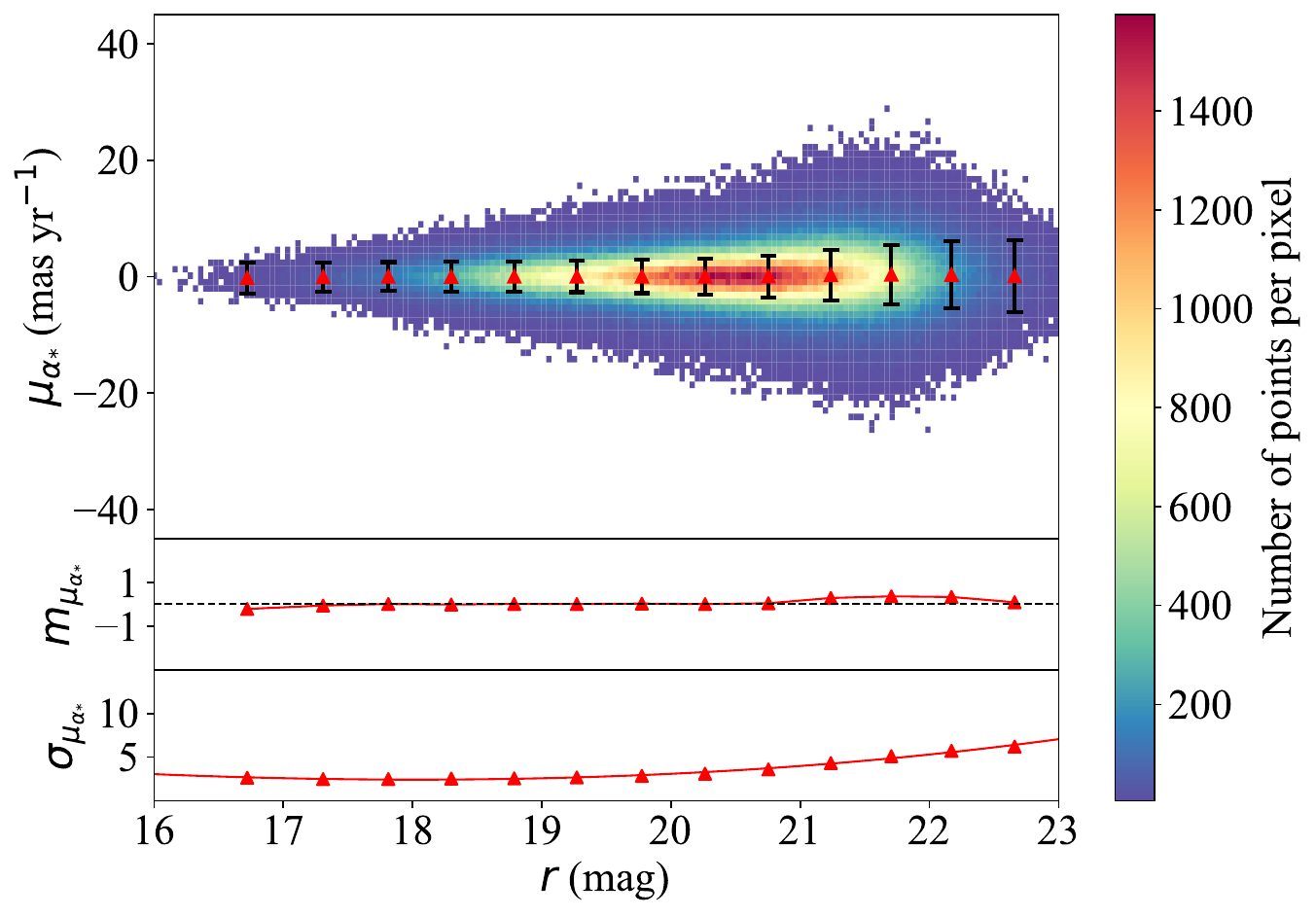}
  \end{subfigure}
  \hfill
  \begin{subfigure}[t]{.48\textwidth}
    \centering
    \includegraphics[width=\linewidth]{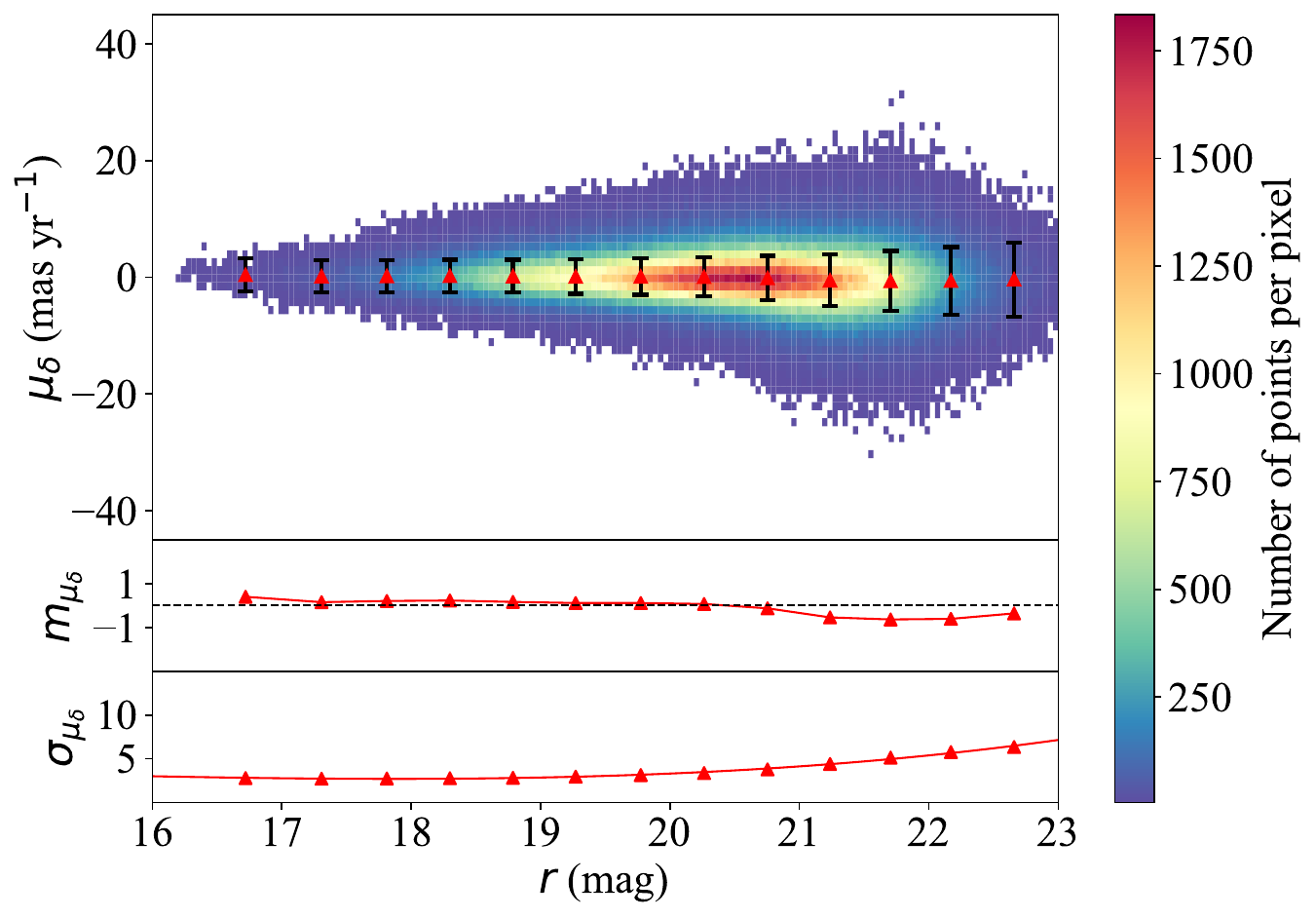}
  \end{subfigure}

  \medskip

  \begin{subfigure}[t]{.48\textwidth}
    \centering
    \includegraphics[width=\linewidth]{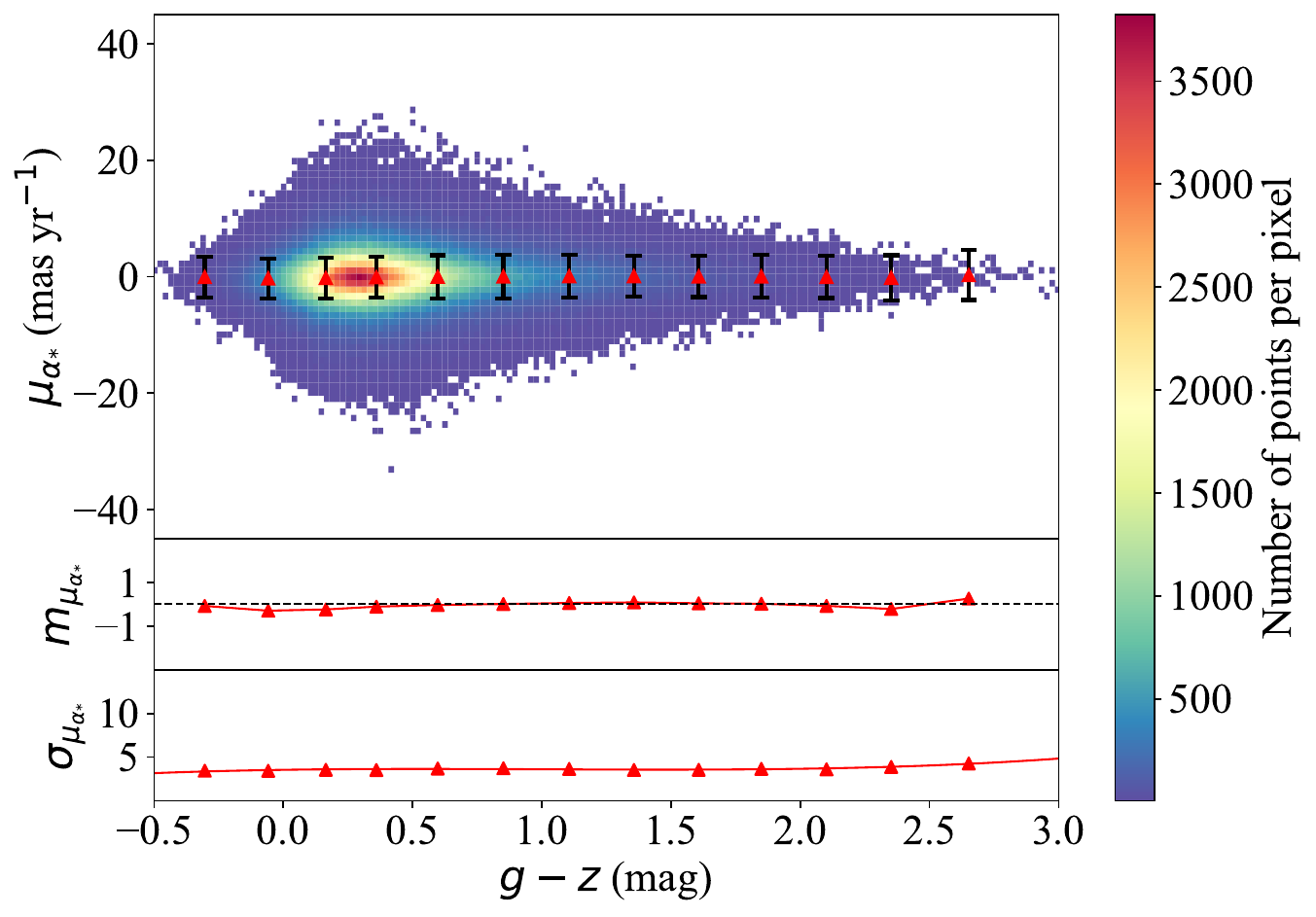}
  \end{subfigure}
  \hfill
  \begin{subfigure}[t]{.48\textwidth}
    \centering
    \includegraphics[width=\linewidth]{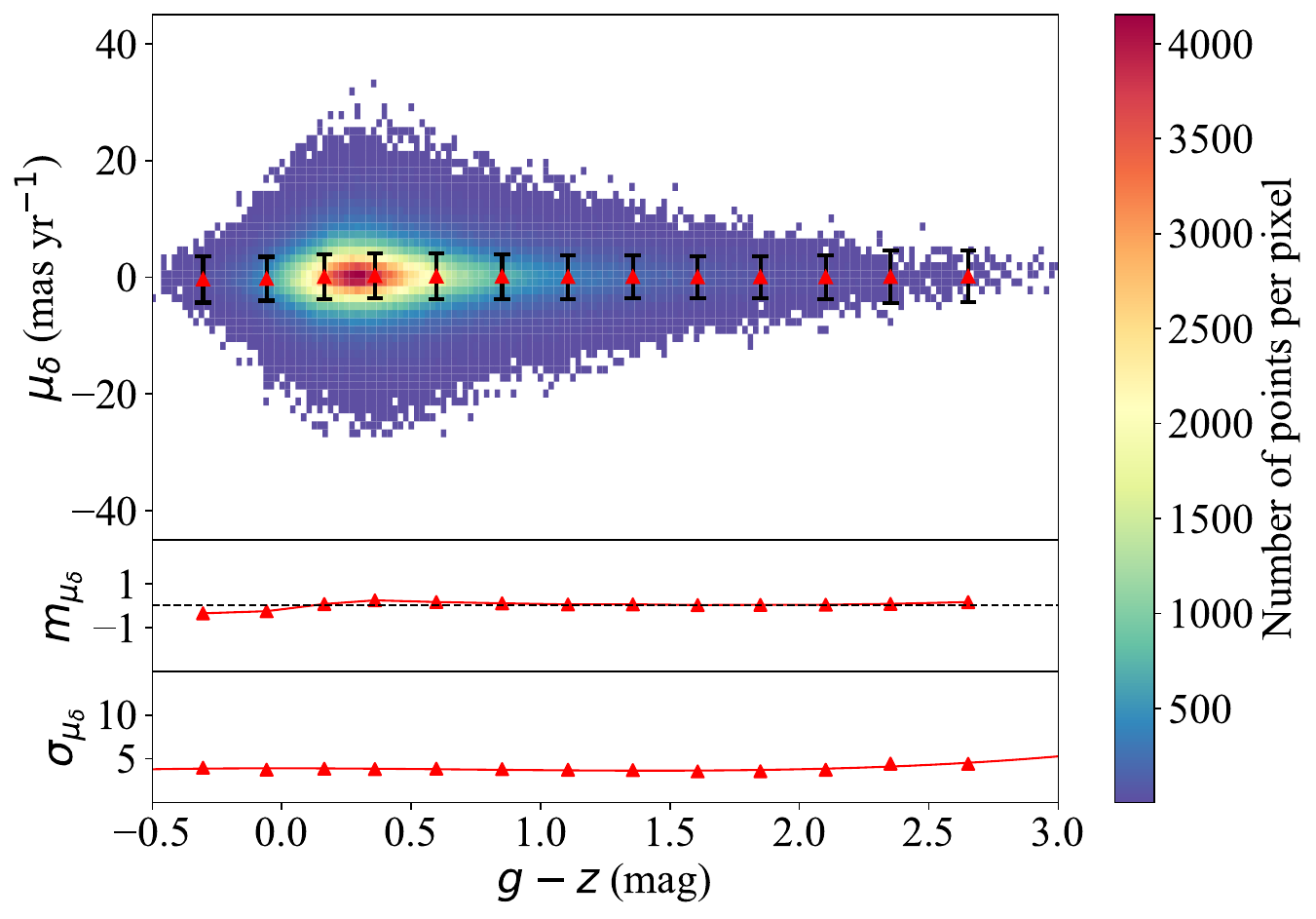}
  \end{subfigure}
  \caption{Proper motions of quasars as functions of color (left subplots) and magnitude (right subplots). The upper and lower subplots separately depict the $\alpha_{*}$ and $\delta$ components. Each subplot encompasses the full dataset of 734 k quasars. The red triangles with error bars represent the median and dispersion of each color or magnitude bin. The median and dispersion as a function of color and magnitude are also plotted in the middle and lower panels.}
  \label{fig:quasar_mag_color_distribution}
\end{figure*}

Beyond the distributions, we further analyze the precision of quasar proper motions with respect to their colors and magnitudes. Figure~\ref{fig:quasar_mag_color_distribution} shows the proper motions of quasars as functions of the $g-z$ color and the $r$-band magnitude. It can be seen that the systematic error exhibits only minimal variation with color. Conversely, redder quasars tend to exhibit more significant random error, a trend potentially skewed by an underrepresentation of quasar samples in the red-color range. Moreover, both systematic and random errors become more pronounced at fainter magnitudes. Specifically, systematic errors surpass 0.5 mas yr$^{-1}$ at $r>21$, while random errors escalate gradually with increasing magnitude, ranging from approximately 2.5 to 5 mas yr$^{-1}$ for objects with $r < 21$. The random error can be as high as around 8 mas yr$^{-1}$ at $r = 23$. The slight systematic biases observed for quasars in Figure~\ref{fig:quasar_mag_color_distribution} arise from their distinct spectral energy distributions compared to galaxies, which can result in slightly different systematic corrections related to magnitude and color.

\subsection{Distant Stars}
\label{sec:4.3}
\begin{figure}[htbp]
    \centering
    \includegraphics[width=0.42\textwidth]{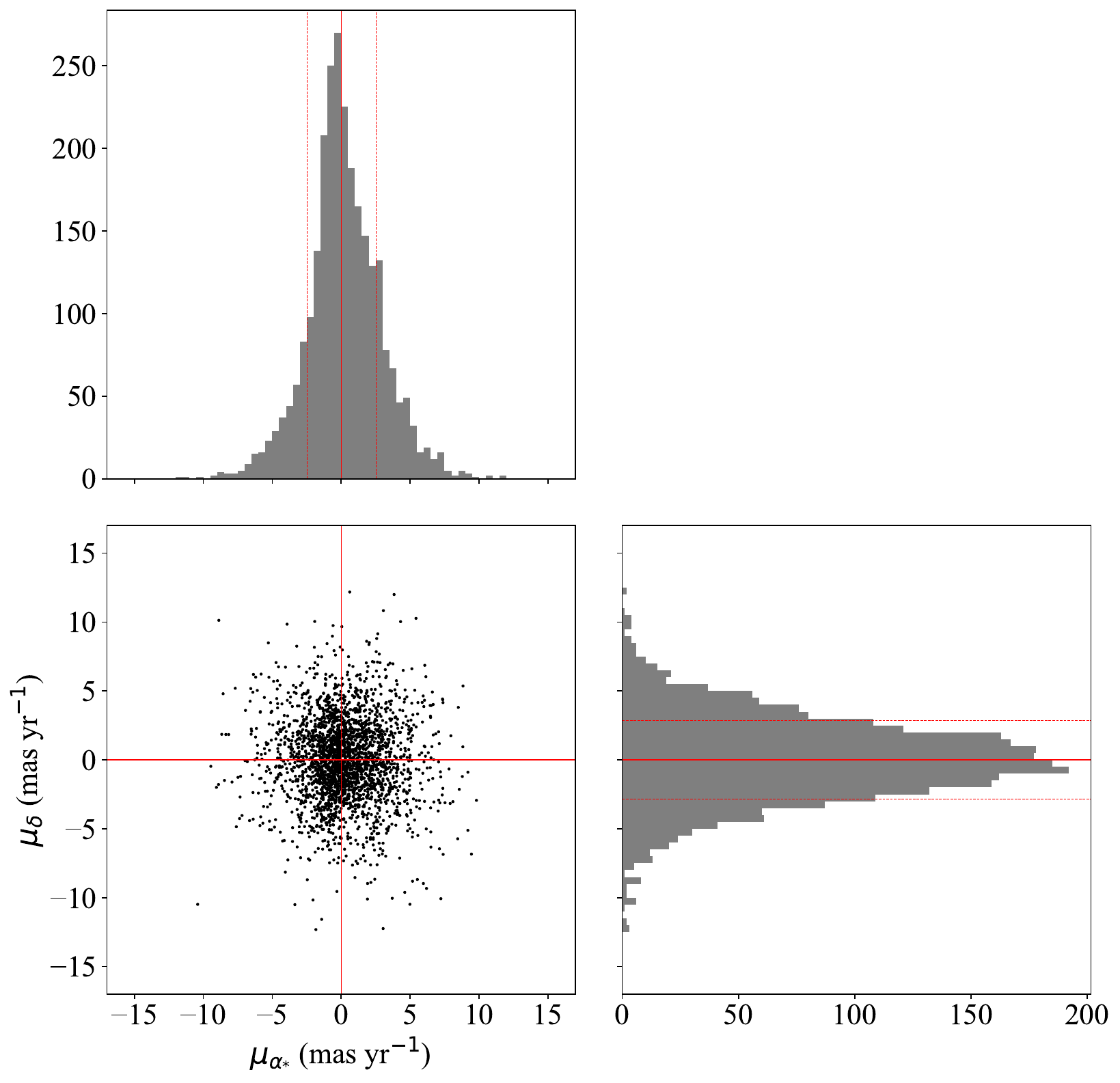}
    \caption{Corrected proper motion distribution of distant halo stars, with solar motion effects accounted for correcting for the solar motion. The top and right subplots present the histograms of $\mu_{\alpha_{*}}$ and $\mu_\delta$, respectively. Red lines denote the median proper motion and corresponding standard deviations for $\alpha_{*}$ and $\delta$.}
    \label{fig:distant_star}
\end{figure}

We employ a total of 2,644 distant (> 20 kpc) halo stars sourced from \citet{2023AJ....165..224Z}, to assess the precision of our proper motion determinations. Although their velocities relative to the Galactic center are small, the influence of solar motion must still be taken into account. For these halo stars even at distances of 20–40 kpc, the reflex motion of the solar causes a noticeable velocity offset (about 1-2 mas yr$^{-1}$, \citet{qiu2021proper}), which can significantly impact the measured proper motions if left uncorrected. By subtracting the solar motion, we isolate the intrinsic motion of the stars themselves, which is crucial for accurately comparing their proper motions. Therefore, a correction for solar motion is implemented, assuming a solar velocity vector of ($U_{\odot}$ , $V_{\odot}$ , $W_{\odot}$) $=$ (11.1, 12.24, 7.25) km s$^{-1}$ \citep{2010MNRAS.403.1829S}, alongside a local standard of rest (LSR) circular velocity of $V_\mathrm{LSR}$ $=$ 220 km s$^{-1}$. The resultant corrected proper motions of these halo stars are shown in Figure \ref{fig:distant_star}. It shows the median values of $\mu_{\alpha_{*}} = $0.03 mas yr$^{-1}$ and $\mu_{\delta} = $0.01 mas yr$^{-1}$ with associated dispersions of 2.52 mas yr $^{-1}$ and 2.86 mas yr$^{-1}$, respectively.

\subsection{Comparison with proper motions from SDSS Stripe 82}
\label{sec:4.5}
\begin{figure*}[htbp]
  \begin{subfigure}[t]{.48\textwidth}
    \centering
    \includegraphics[width=\linewidth]{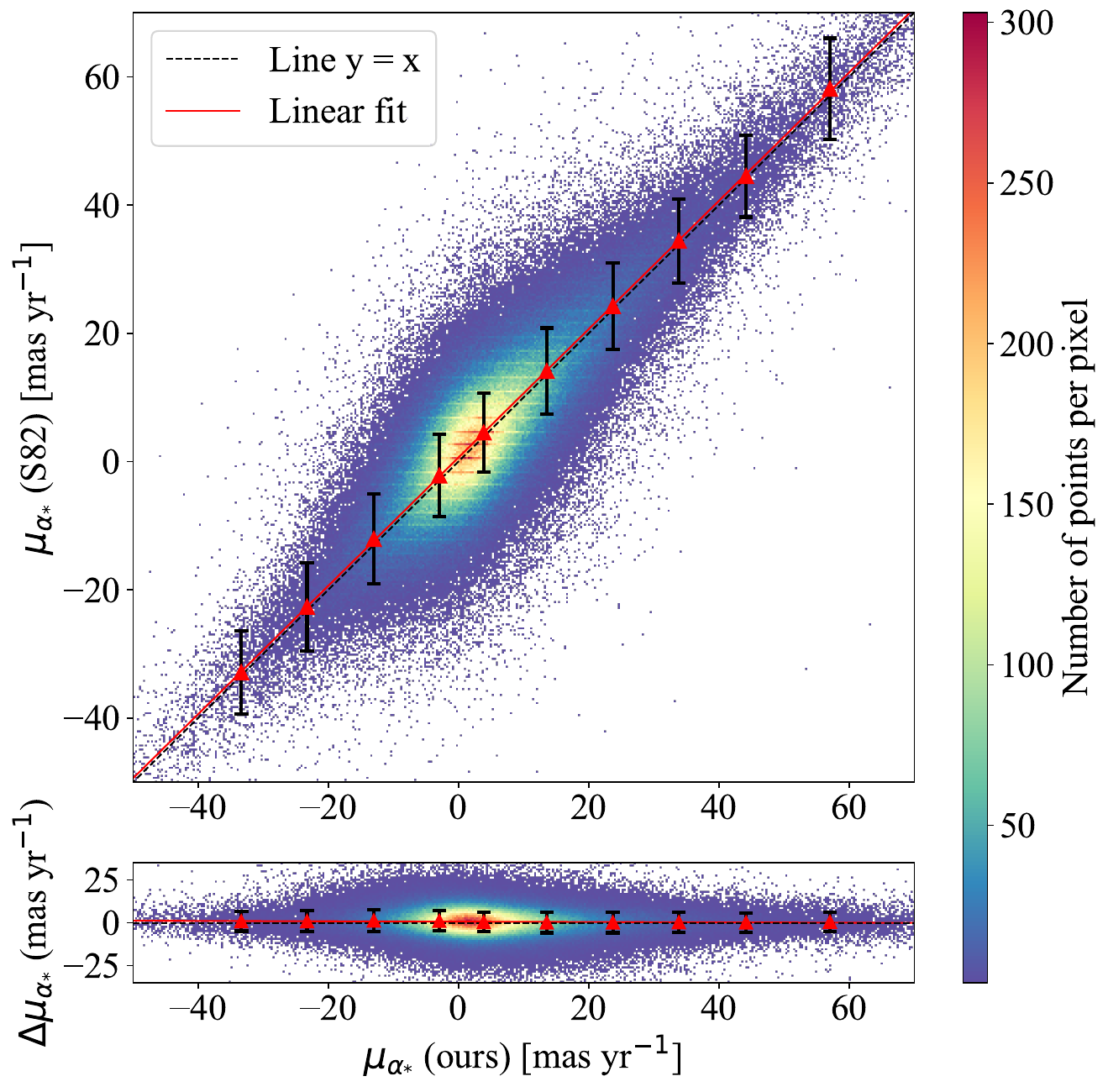}
  \end{subfigure}
  \hfill
  \begin{subfigure}[t]{.48\textwidth}
    \centering
    \includegraphics[width=\linewidth]{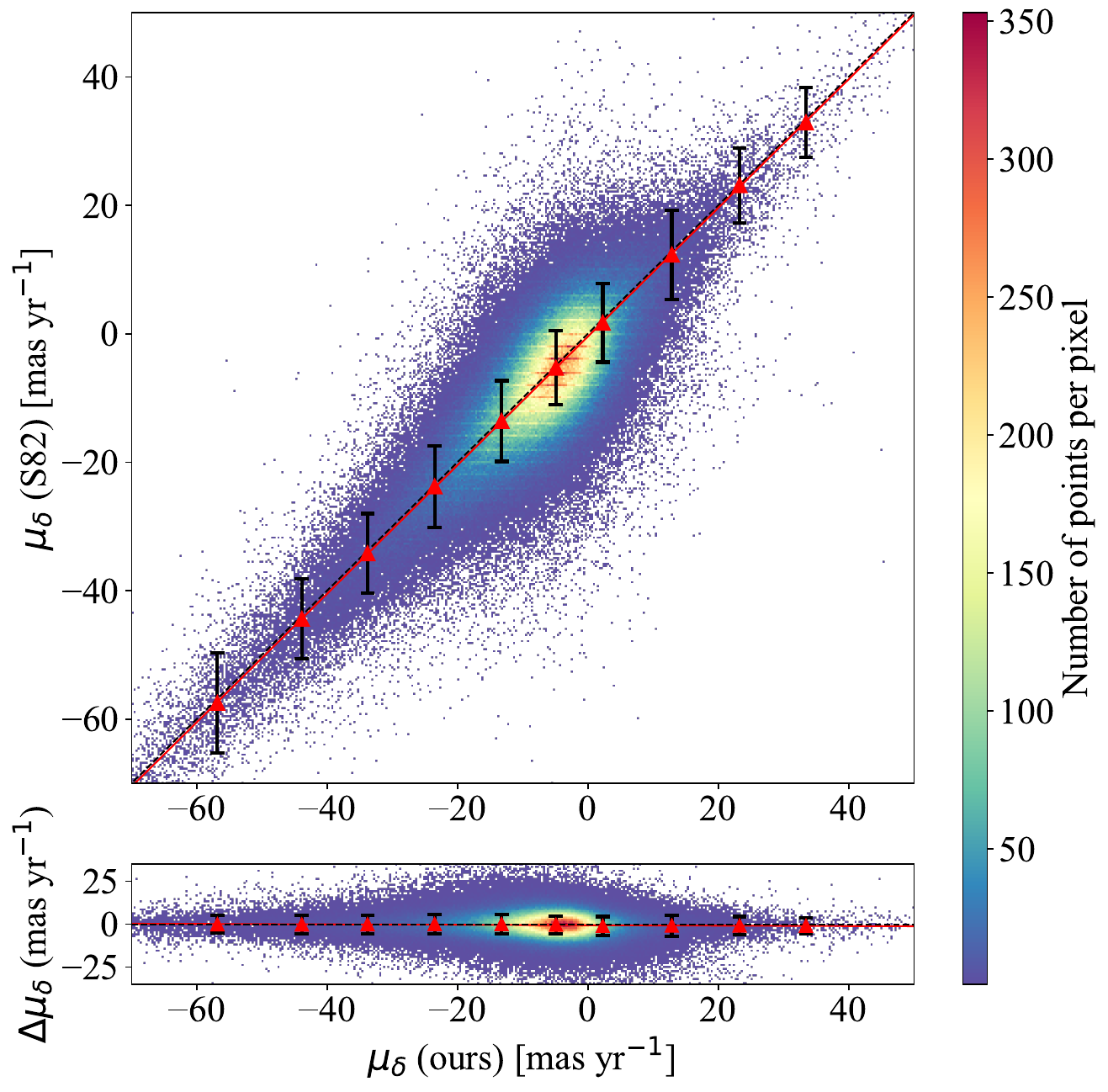}
  \end{subfigure}
  \caption{Comparison of our proper motions with those from SDSS S82 for $\mu_{\alpha_{*}}$ (left) and $\mu_{\delta}$ (right). The red triangles with error bars represent the median and dispersion of each proper motion bin. Red solid lines present the linear relationship of our and S82 measurements, and the black dashed lines in top and bottom panels mark $y=x$ and $y=0$, respectively.}
  \label{fig:stripe82}
\end{figure*}

\citet{bramich2008light} released a proper motion catalog containing nearly 4 million objects in the SDSS Stripe 82 (hereafter S82). Through a cross-match within a radius of 1 arcsecond and applying simple constraints on the proper motion and photometric errors of S82, we obtained a sample of approximately 1 million stars. Figure \ref{fig:stripe82} illustrates the comparison of our proper motions of these chosen stars with those of S82. A rigorous linear relationship between the two sets of proper motion measurements can be seen from this figure, although there is a considerable dispersion. This scatter is largely attributed to the astrometric inaccuracies in S82 as reported in \citet{bramich2008light}, particularly for sources fainter than $r=20$.

\section{Our proper motion catalog and an application example} \label{sec:5}
\subsection{Our proper motion catalog and its limitations}
Our DESI-SDSS proper motion catalog provides the proper motion measurements for 223.7 million non-Gaia objects. The catalog can be accessed at \url{https://doi.org/10.57760/sciencedb.07990}. We use quasars to estimate the proper motion random error for point sources:
\begin{equation}
    \sigma_{\mu_{\alpha_{*},\delta}} (i,j) = f_{i,j}(r, g-z),
    \label{eql:3}
\end{equation}
where $i$ and $j$ represent the RA and DEC coordinates of a given pixel, and $f(r, g-z)$ denotes a series of two-dimensional polynomial models fitted to the proper motion systematic errors ($\mu_{\alpha_{*}}$ and $\mu_\delta$), accounting for variations in magnitude and color. For a given pixel at coordinates ($\alpha_{*i}, \delta_j$), the corresponding function $f_{i,j}(r, g-z)$ is derived by fitting quasar samples located within a 5$^{\circ}$ radius around that pixel.

Table \ref{tab:catalog} in Appendix \ref{sec:cat1} outlines the content of our catalog. In addition to proper motion and its error, it encompasses fundamental details extracted from both the SDSS and DESI databases, such as source positions, magnitudes, and object classifications. For enhanced user accessibility, we have incorporated the Gaia proper motion data as well, which come from the released data of DR2 \footnote{\url{https://www.legacysurvey.org/dr10/description/\#astrometry}}.

It should be noted that our proper motion catalog also has several limitations:
\begin{itemize}
    \item Our analysis focused exclusively on the proper motion of objects not covered by Gaia, as the current DESI LS catalog lacks positional data for sources already detected by Gaia.
    \item A small fraction of pixels ($\sim1\%$) spans multiple observation regions, leading to multiple epoch distributions within a single pixel. These pixels are flagged (FLAG$=$1) in the final catalog to alert users to exercise caution when using the associated proper motions.
    \item Employing galaxies from the DESI Legacy Survey as the reference frame is susceptible to distortive effects induced by DCR. This issue is further compounded by the reliance on co-added positional information from multiple exposures inherent to the current DESI catalog, thereby increasing the complexity of the analysis.
    \item \textit{The Tractor} simultaneously fits source models to individual exposures from multiple epochs, minimizing the residuals to determine the best-fit position for each source. However, accurately deriving the epoch for the fitted position is challenging. The current DESI epoch is calculated as the average of the MJD values from all relevant exposures, which may introduce uncertainties in the measurement of proper motions.
    \item The systematic biases observed in quasars stem from their unique spectral energy distributions, which differ from those of galaxies. These differences result in varying systematic corrections, particularly concerning magnitude and color.
    \item The proper motion uncertainties, represented by PMRA\_ERR and PMDEC\_ERR, are provided as statistical estimates and do not reflect measurements derived from individual targets. These uncertainties are primarily an indication of the goodness of the proper motions and should not be used for scientific investigations, such as measuring the intrinsic velocity dispersion of stellar systems.
    \item Targets observed in the SGC feature a shorter time interval between observations, leading to proper motion measurement errors that are two to three times greater compared to those in other regions.
    \item Both SDSS and DESI supply morphological classifications for the objects as provided in our catalog. Users can incorporate additional information such as color to achieve a more refined physical classification.
\end{itemize}

\subsection{Applying our catalog to estimate the proper motions of star clusters}

Stars within a star cluster exhibit a collective motion, enabling us to leverage clusters with precisely known proper motions to validate our proper motion measurements and refine them using fainter cluster members. This study employs star clusters documented in the Milky Way Star Clusters Catalog \citep[MWSC;][]{2012A&A...543A.156K,2013A&A...558A..53K,2014A&A...568A..51S,2015A&A...581A..39S}. Only 15 star clusters have sufficient number of members in the DESI-SDSS footprint. 

\begin{figure*}[htbp]

  \begin{subfigure}[t]{.33\textwidth}
    \centering
    \includegraphics[width=\linewidth]{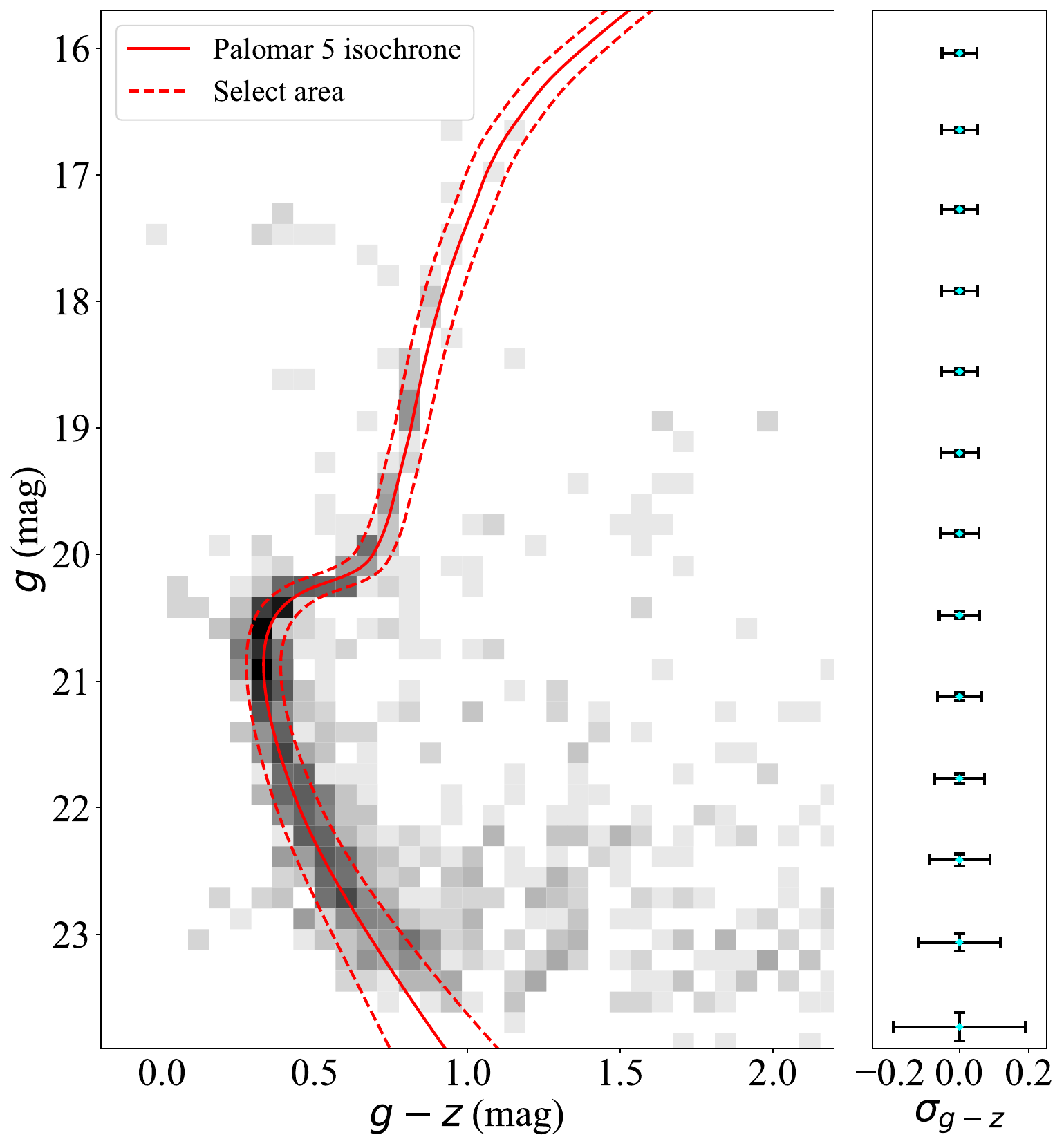}
  \end{subfigure}
  \hfill
  \begin{subfigure}[t]{.26\textwidth}
    \centering
    \includegraphics[width=\linewidth]{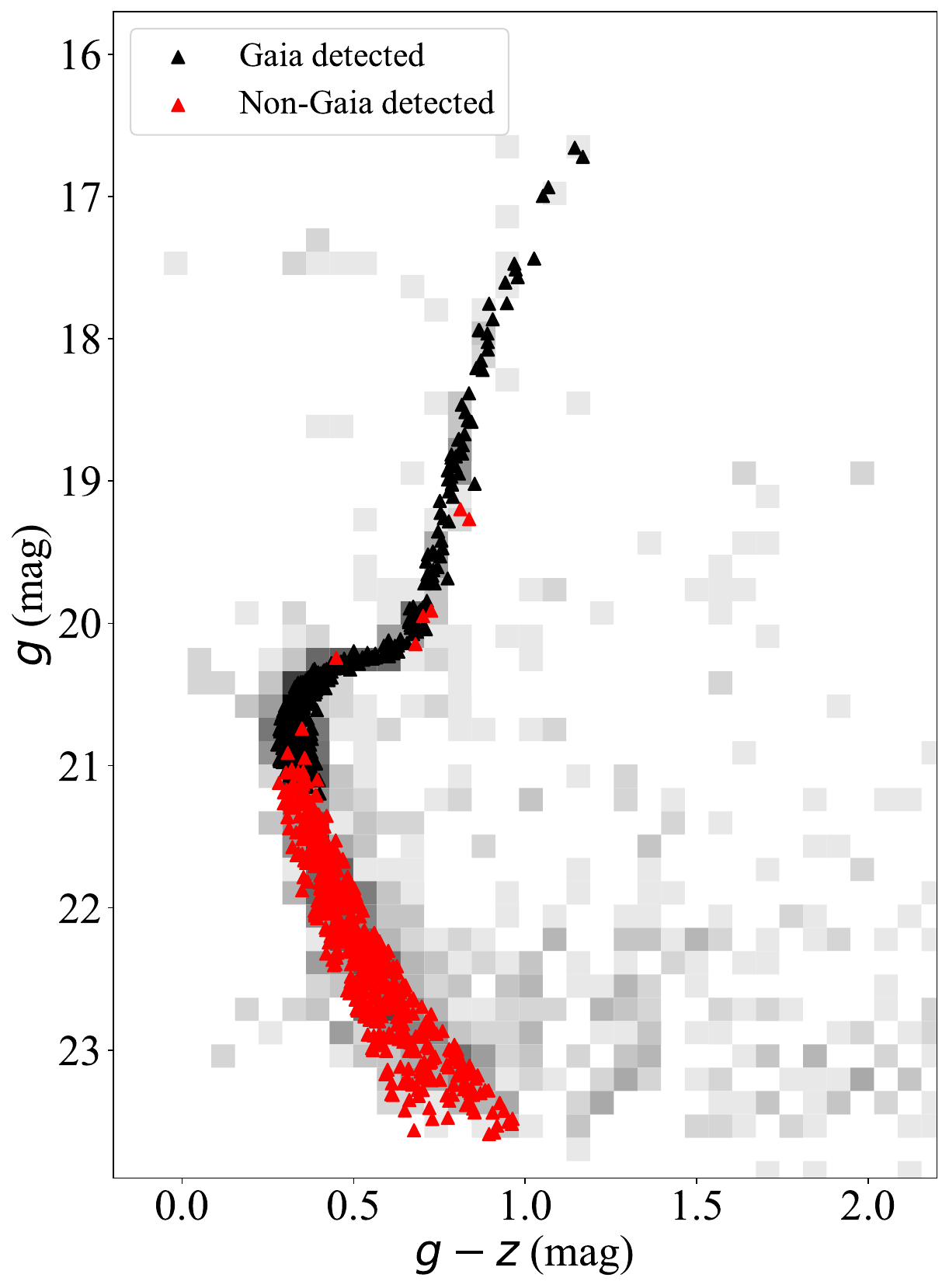}
  \end{subfigure}
  \hfill
  \begin{subfigure}[t]{.36\textwidth}
    \centering
    \includegraphics[width=\linewidth]{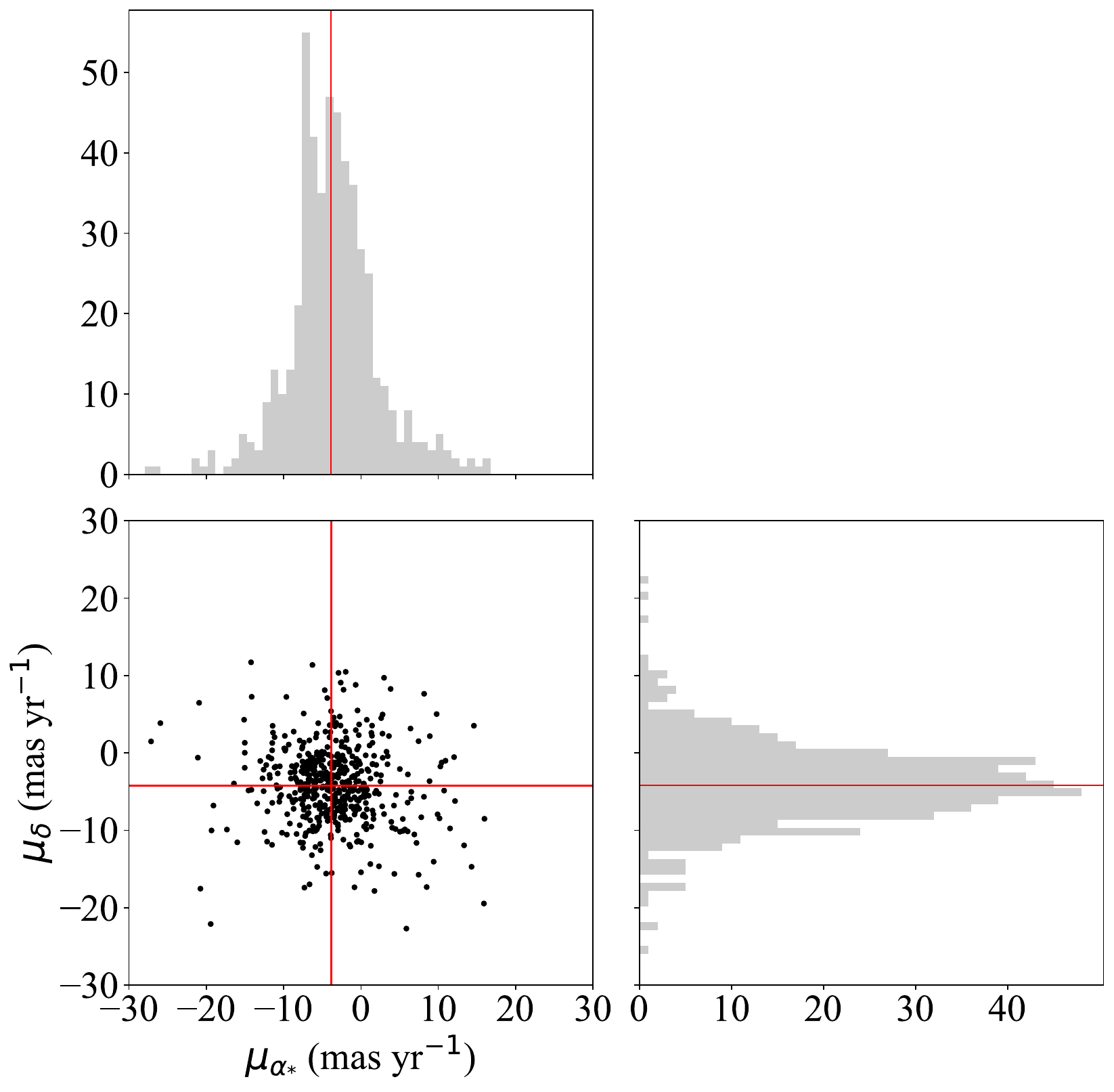}
  \end{subfigure}

  \caption{Left: color-magnitude diagram for Palomar 5, featuring all stars within the cluster radius of 0.14{\arcdeg}. The solid red curve delineates the theoretical isochrone, while the dashed red envelopes enclosing it symbolize a combined 1$\sigma$ photometric uncertainty coupled with an intrinsic color error of 0.05 mag. To the right of the CMD, cyan dots accompanied by error bars illustrate the magnitude and color measurement uncertainties. The region enclosed by the dashed lines is used to select member stars. Middle: selected Gaia (black triangles) and non-Gaia (red triangles) member stars within the CMD. Right: proper motion distribution of non-Gaia member stars. The median values for $\alpha_{*}$ and $\delta$ are marked with solid red lines. The top and right subplots present the distributions of $\mu_{\alpha_{*}}$ and $\mu_\delta$, respectively.}
  
  \label{fig:clusters}
\end{figure*}

For each cluster, we select cluster members as those stars within the cluster radius and located close to the theoretical isochrone on the color-magnitude diagram (CMD). The cluster radius and the stellar population parameters (e.g. age, metallicity and distance modulus) come from the MWSC catalog. Theoretical isochrones are obtained from the PARSEC\footnote{\url{http://stev.oapd.inaf.it/cgi-bin/cmd}} evolutionary tracks \citep{bressan2012parsec}. The left panel of Figure \ref{fig:clusters} presents the CMD overlaid with the theoretical isochrone for Palomar 5 as an example. Stars belonging to a cluster exhibit inherent scatter in their positioning along the isochrone on the CMD. Consequently, we identify potential cluster members as those stars situated within a radius of 0.14{\arcdeg} of the cluster center, along with those that fall within a band encompassing 1$\sigma$ photometric uncertainty augmented by an additional intrinsic dispersion of 0.05 mag around the isochrone path (dashed curves in Figure \ref{fig:clusters}). The right panel of Figure \ref{fig:clusters} illustrates the proper motion distribution for the cluster members that are not part of the Gaia dataset. The mean proper motion of the cluster is estimated as $\mu_{\alpha_{*}}= -$3.86 mas yr$^{-1}$ and $\mu_\delta= -$4.20 mas yr$^{-1}$. The corresponding errors are 0.22 mas yr$^{-1}$ for $\alpha_{*}$ and 0.21 mas yr$^{-1}$ for $\delta$. 

\begin{figure*}[htbp]
  \begin{subfigure}[t]{.48\textwidth}
    \centering
    \includegraphics[width=\linewidth]{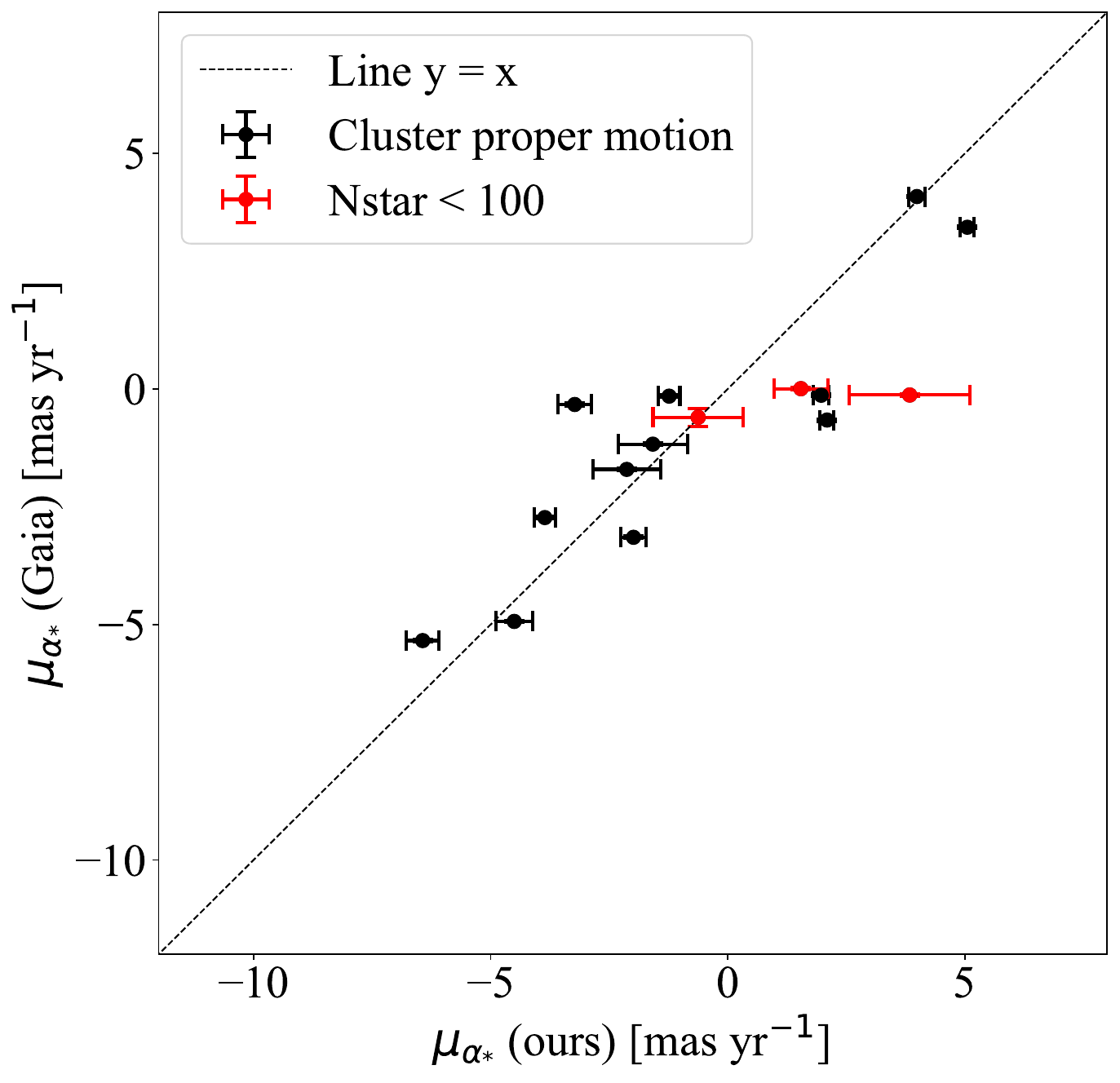}
  \end{subfigure}
  \hfill
  \begin{subfigure}[t]{.48\textwidth}
    \centering
    \includegraphics[width=\linewidth]{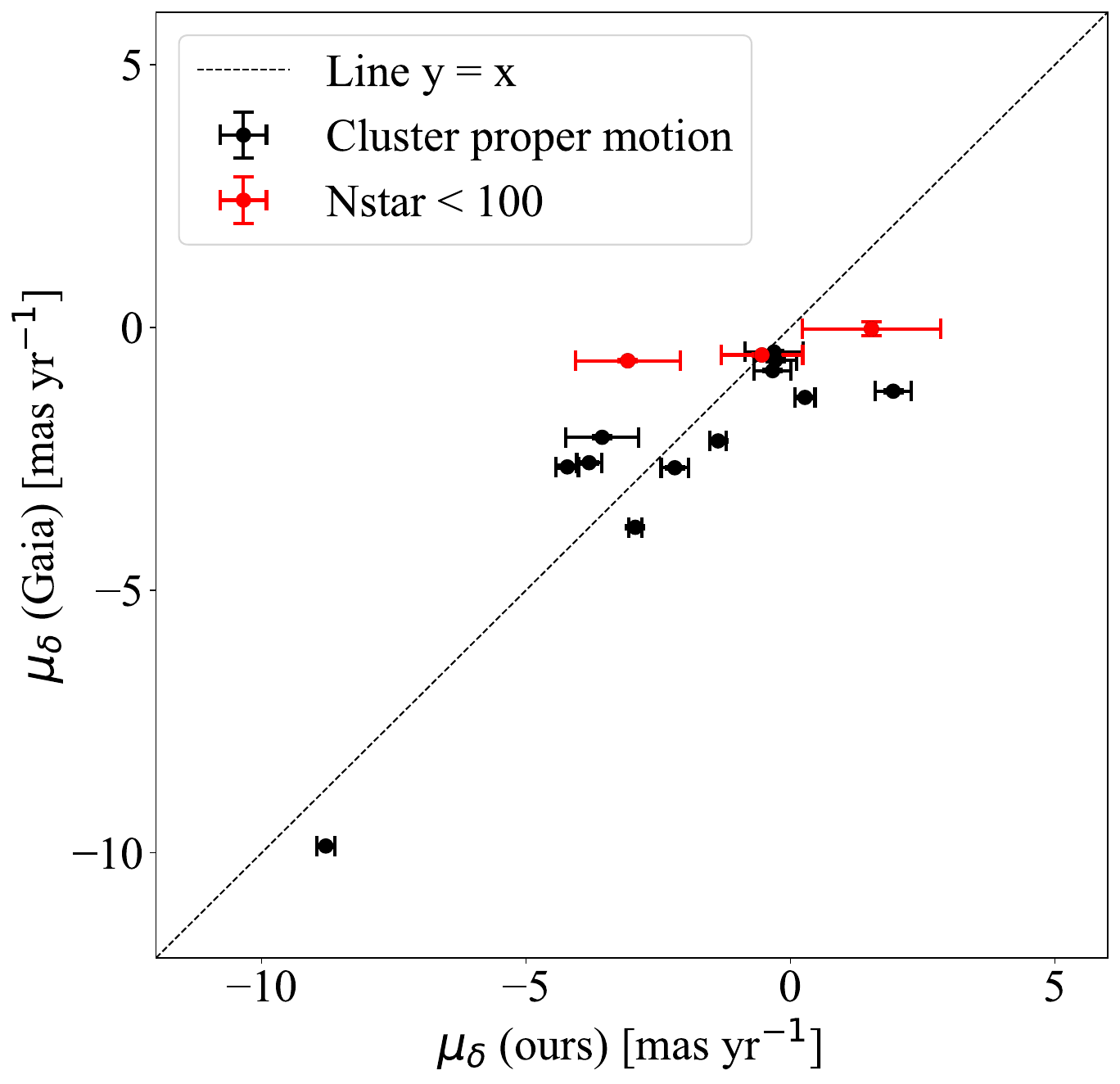}
  \end{subfigure}
  
  \caption{Comparison of proper motions for 15 star clusters derived from both our and Gaia catalogs. The left panel focuses on the comparison of $\mu_{\alpha_{*}}$. The right panel presents the corresponding comparison for $\mu_\delta$. The black dashed line displays $y=x$. Data points marked in red represent clusters with the number of DESI-SDSS or Gaia member stars less than 100.}

  \label{fig:clustergaia}
\end{figure*}

Figure \ref{fig:clustergaia} compares the proper motions for 15 star clusters derived from our measurements and those from \citet{vasiliev2021gaia}. Table~\ref{tab:cluster} in Appendix \ref{sec:cat2} provides the basic information and proper motions for all these clusters. The data in Figure \ref{fig:clustergaia} show a general correlation between the two measurements, though some scatter exists. Notably, the clusters displaying the most significant discrepancies between our SDSS/DESI catalog and Gaia data tend to have insufficient members (as marked in red in this figure). It is also worth mentioning that our method for member selection in calculating proper motions differs from that of \citet{vasiliev2021gaia}, which may contribute to the observed discrepancies.

\section{Summary}
\label{sec:6}
We have successfully established the DESI-SDSS proper motion catalog through the construction of a reference frame utilizing galaxies, with subsequent corrections for position-, color-, and magnitude-dependent astrometric discrepancies. Despite the relatively lower astrometric precision of ground-based surveys such as SDSS and DESI in contrast to space-based counterparts, the extensive temporal baseline of about 13 years partly mitigates this deficiency. Our proper motion accuracy has been validated through multiple tests involving galaxies, quasars, remote stars, and cross-referencing with SDSS Stripe 82 data. 

Based on an analysis of 734,012 quasars, the systematic errors in our proper motion measurements were quantified at 0.06 mas yr$^{-1}$ in $\alpha_{*}$ and 0.15 mas yr$^{-1}$ in $\delta$, with random errors of 3.23 mas yr$^{-1}$ and 3.57 mas yr$^{-1}$ in $\alpha_{*}$ and $\delta$, respectively. The proper motion of stars and galaxies reveals a systematic difference in the proper motion errors between NGC and SGC due to different time baselines. The distant halo stars give a precision better than 3.0 mas yr$^{-1}$. A comparative study with SDSS Stripe 82 using near 1 million stars confirms a favorable agreement, reinforcing the reliability of our catalog. 

As a practical demonstration, we applied our catalog to calculate the proper motions for 15 star clusters. The agreement between the proper motions derived from Gaia and those from our DESI-SDSS catalog demonstrates a strong consistency. Thus, our DESI-SDSS proper motion catalog stands as a valuable resource, effectively bridging the sensitivity gap left by Gaia at the faint magnitude end. Coupled with extensive DESI imaging data, this catalog provides relatively accurate proper motions for stars fainter than those in the Gaia dataset. These data can enhance the identification of member features in existing halo substructures, stellar streams, and satellite galaxies. They also facilitate the detection of fainter structures in these objects, the discovery of new substructures, and the identification of faint hypervelocity stars. More accurate proper motions can be derived in the future once single-epoch positions from individual DESI exposures become available.

\acknowledgments
The authors acknowledge the supports from the Beijing Municipal Natural Science Foundation (grant No. 1222028), the National Key R\&D Program of China (grant Nos. 2022YFA1602902, 2023YFA1607800, 2023YFA1607804, 2023YFA1608100, and 2023YFF0714800), and the National Natural Science Foundation of China (NSFC; grant Nos. 12120101003, 12373010, 12173051, and 12233008). The authors also acknowledge the science research grants from the China Manned Space Project with Nos. CMS-CSST-2021-A02 and CMS-CSST-2021-A04 and the Strategic Priority Research Program of the Chinese Academy of Sciences with Grant Nos. XDB0550100 and XDB0550000.

The DESI Legacy Imaging Surveys consist of three individual and complementary projects: the Dark Energy Camera Legacy Survey (DECaLS), the Beijing-Arizona Sky Survey (BASS), and the Mayall z-band Legacy Survey (MzLS). DECaLS, BASS and MzLS together include data obtained, respectively, at the Blanco telescope, Cerro Tololo Inter-American Observatory, NSF’s NOIRLab; the Bok telescope, Steward Observatory, University of Arizona; and the Mayall telescope, Kitt Peak National Observatory, NOIRLab. NOIRLab is operated by the Association of Universities for Research in Astronomy (AURA) under a cooperative agreement with the National Science Foundation. Pipeline processing and analyses of the data were supported by NOIRLab and the Lawrence Berkeley National Laboratory (LBNL). Legacy Surveys also uses data products from the Near-Earth Object Wide-field Infrared Survey Explorer (NEOWISE), a project of the Jet Propulsion Laboratory/California Institute of Technology, funded by the National Aeronautics and Space Administration. Legacy Surveys was supported by: the Director, Office of Science, Office of High Energy Physics of the U.S. Department of Energy; the National Energy Research Scientific Computing Center, a DOE Office of Science User Facility; the U.S. National Science Foundation, Division of Astronomical Sciences; the National Astronomical Observatories of China, the Chinese Academy of Sciences and the Chinese National Natural Science Foundation. LBNL is managed by the Regents of the University of California under contract to the U.S. Department of Energy. The complete acknowledgements can be found at \url{https://www.legacysurvey.org/acknowledgment/}.

Funding for SDSS-III has been provided by the Alfred P. Sloan Foundation, the Participating Institutions, the National Science Foundation, and the U.S. Department of Energy Office of Science. The SDSS-III web site is \url{http://www.sdss3.org/}. SDSS-III is managed by the Astrophysical Research Consortium for the Participating Institutions of the SDSS-III Collaboration including the University of Arizona, the Brazilian Participation Group, Brookhaven National Laboratory, Carnegie Mellon University, University of Florida, the French Participation Group, the German Participation Group, Harvard University, the Instituto de Astrofisica de Canarias, the Michigan State/Notre Dame/JINA Participation Group, Johns Hopkins University, Lawrence Berkeley National Laboratory, Max Planck Institute for Astrophysics, Max Planck Institute for Extraterrestrial Physics, New Mexico State University, New York University, Ohio State University, Pennsylvania State University, University of Portsmouth, Princeton University, the Spanish Participation Group, University of Tokyo, University of Utah, Vanderbilt University, University of Virginia, University of Washington, and Yale University.

This work presents results from the European Space Agency (ESA) space mission Gaia. Gaia data are being processed by the Gaia Data Processing and Analysis Consortium (DPAC). Funding for the DPAC is provided by national institutions, in particular the institutions participating in the Gaia MultiLateral Agreement (MLA). The Gaia mission website is \url{https://www.cosmos.esa.int/gaia}. The Gaia archive website is \url{https://archives.esac.esa.int/gaia}.

\bibliography{pm_desi}{}
\bibliographystyle{aasjournal}

\appendix
\section{Our proper motion catalog} \label{sec:cat1}
Table \ref{tab:catalog} summarizes the basic information of our proper motion catalog. It comprises photometric parameters originating from the DESI and SDSS photometric datasets, such as celestial coordinates, magnitudes, and morphological classifications. The catalog provides the proper motions and corresponding errors for objects not included in Gaia, which are derived from our comparative analysis between the SDSS and DESI astrometric measurements. It also contains Gaia objects, whose proper motions are inherited from the Gaia DR2. 
\begin{table}[htbp!]
    \centering
    {
    \begin{threeparttable}
    \centering
    \caption{Our proper motion catalog.}
    \setlength{\tabcolsep}{6pt}
    \begin{tabular}{l l c c l}
    \hline
    \hline
    \addlinespace
    Column & Unit & Format & Description \\
    \addlinespace
    \hline
    \addlinespace
    objID         &               & Long     & SDSS unique object identifier \\
    ID            &               & Long     & Object ID in DESI \\
    RA            & degree        & Double   & Right ascension in J2000 from DESI \\
    RA\_IVAR      & 1/degree$^2$  & Float    & Inverse variance of RA \\
    DEC           & degree        & Double   & Declination in J2000 from DESI \\
    DEC\_IVAR     & 1/degree$^2$  & Float    & Inverse variance of DEC \\
    PM\_SOURCE    &               & String   & Source of proper motion (Gaia or DESI) \\
    FLAG          &               & Integer  & Indicator of observation time variation within the source’s pixel \\
                  &               &          & (0 indicates no issues, while 1 suggests caution is needed) \\
    PMRA          & mas yr$^{-1}$ & Double   & Proper motion for $\alpha_{*}$ \\
    PMDEC         & mas yr$^{-1}$ & Double   & Proper motion for $\delta$ \\
    PMRA\_ERR     & mas yr$^{-1}$ & Double   & proper motion error for $\alpha_{*}$ \\
    PMDEC\_ERR    & mas yr$^{-1}$ & Double   & proper motion error for $\delta$ \\
    MJD\_SDSS     & days          & Integer  & Julian date of the SDSS observation \\
    MJD\_DESI     & days          & Double   & mean Julian date of the DESI observations \\
    TIME          & year          & Double   & Time interval between SDSS and DESI observation \\
    TYPE\_SDSS    &               & Short    & Type of object (3=galaxy, 6=star) classified by SDSS \\
    TYPE\_DESI\tnote{a} &         & String   & Type of object classified by DESI \\
    MAG\_G        & mag           & Float    & DESI $g$-band magnitude \\
    MAG\_R        & mag           & Float    & DESI $r$-band magnitude \\
    MAG\_Z        & mag           & Float    & DESI $z$-band magnitude \\
    MAGERR\_G     & mag           & Float    & DESI $g$-band magnitude error \\
    MAGERR\_R     & mag           & Float    & DESI $r$-band magnitude error  \\
    MAGERR\_Z     & mag           & Float    & DESI $z$-band magnitude error  \\
    \addlinespace
    \hline
    \end{tabular}
    \begin{tablenotes}
         \item [a] See more details at \url{https://www.legacysurvey.org/dr10/description/\#morphological-classification}.
    \end{tablenotes}
    \label{tab:catalog}
    \end{threeparttable}
    }
\end{table}

\section{Proper motions for 15 star clusters} \label{sec:cat2}
Table \ref{tab:cluster} lists the proper motions estimated with both our catalog and the Gaia data. \label{sec:cat1}
\begin{sidewaystable}[htbp]
    \centering
    \resizebox{\textwidth}{!}
    {
    \begin{threeparttable}
    \centering
    \caption{Proper motion and other parameters for 15 star clusters used in this work}
    \setlength{\tabcolsep}{6pt}
    \begin{tabular}{cccccccccccccccccc}
    \hline
    \hline
    \addlinespace
    NAME & RA & DEC & EBV & R & D & [Fe/H] & AGE & $\mu_{\alpha_{*}}$\tnote{a} & $\sigma_{\mu_{\alpha_{*}}}$\tnote{a} & $\mu_{\delta}$\tnote{a} & $\sigma_{\mu_{\delta}}$\tnote{a} & num\tnote{a} & $\mu_{\alpha_{*}}$\tnote{b} & $\sigma_{\mu_{\alpha_{*}}}$\tnote{b} & $\mu_{\delta}$\tnote{b} & $\sigma_{\mu_{\delta}}$\tnote{b} \\
    \addlinespace
    (1) & (2) & (3) & (4) & (5) & (6) & (7) & (8) & & & & & & & & & & \\
    \addlinespace
    \hline
    \addlinespace
    Koposov 2 & 119.571 & 26.25 & 0.021 & 0.07 & 16.725 & -2.155 & 10.1 & -0.624 & 0.952 & 1.531 & 1.306 & 43 & -0.601 & 0.189 & -0.025 & 0.129 \\ 
    \addlinespace
    NGC 2419 & 114.535 & 38.882 & 0.083 & 0.12 & 19.61 & -2.05 & 10.1 & 1.543 & 0.565 & -0.540 & 0.773 & 96 & 0.007 & 0.028 & -0.523 & 0.026 \\ 
    \addlinespace
    NGC 4147 & 182.526 & 18.543 & 0.021 & 0.06 & 16.435 & -1.645 & 10.1 & -2.130 & 0.713 & -3.561 & 0.683 & 174 & -1.707 & 0.027 & -2.09 & 0.027 \\ 
    \addlinespace
    NGC 5024 & 198.23 & 18.168 & 0.021 & 0.25 & 16.27 & -1.75 & 10.1 & 1.971 & 0.164 & 0.275 & 0.189 & 1546 & -0.133 & 0.024 & -1.331 & 0.024 \\ 
    \addlinespace
    NGC 5053 & 199.113 & 17.7 & 0.01 & 0.15 & 16.206 & -2.15 & 10.1 & -3.228 & 0.352 & 1.944 & 0.343 & 543 & -0.329 & 0.025 & -1.213 & 0.025 \\ 
    \addlinespace
    NGC 5272 & 205.548 & 28.377 & 0.01 & 0.44 & 15.045 & -1.345 & 10.1 & -1.236 & 0.228 & -2.187 & 0.259 & 1506 & -0.152 & 0.023 & -2.67 & 0.022 \\ 
    \addlinespace
    NGC 5466 & 211.38 & 28.534 & 0 & 0.205 & 16.021 & -1.745 & 10.1 & -6.434 & 0.345 & -0.337 & 0.349 & 751 & -5.342 & 0.025 & -0.822 & 0.024 \\ 
    \addlinespace
    NGC 5904 & 229.638 & 2.081 & 0.029 & 0.44 & 14.385 & -1.145 & 10.1 & 3.988 & 0.173 & -8.788 & 0.170 & 1900 & 4.086 & 0.023 & -9.87 & 0.023 \\ 
    \addlinespace
    NGC 6205 & 250.422 & 36.46 & 0.021 & 0.47 & 14.265 & -1.445 & 10.1 & -1.987 & 0.264 & -3.808 & 0.233 & 1403 & -3.149 & 0.023 & -2.574 & 0.023 \\ 
    \addlinespace
    NGC 6229 & 251.745 & 47.528 & 0.01 & 0.165 & 17.425 & -1.145 & 10.1 & -1.580 & 0.728 & -0.311 & 0.545 & 148 & -1.171 & 0.026 & -0.467 & 0.027 \\ 
    \addlinespace
    NGC 6341 & 259.281 & 43.136 & 0.021 & 0.405 & 14.6 & -2.05 & 10.1 & -4.503 & 0.393 & -0.285 & 0.403 & 880 & -4.935 & 0.024 & -0.625 & 0.024 \\ 
    \addlinespace
    NGC 7006 & 315.372 & 16.187 & 0.052 & 0.125 & 18.09 & -1.52 & 10.1 & 3.838 & 1.273 & -3.077 & 0.993 & 25 & -0.128 & 0.027 & -0.633 & 0.027 \\ 
    \addlinespace
    NGC 7078 & 322.493 & 12.167 & 0.104 & 0.35 & 15.117 & -2.05 & 10.1 & 2.094 & 0.146 & -2.935 & 0.122 & 1468 & -0.659 & 0.024 & -3.803 & 0.024 \\ 
    \addlinespace
    NGC 7089 & 323.363 & -0.823 & 0.062 & 0.26 & 15.325 & -1.545 & 10.1 & 5.053 & 0.146 & -1.369 & 0.153 & 1382 & 3.435 & 0.025 & -2.159 & 0.024 \\ 
    \addlinespace
    Palomar 5 & 229.022 & -0.112 & 0.031 & 0.14 & 16.837 & -1.245 & 10.1 & -3.858 & 0.220 & -4.220 & 0.210 & 473 & -2.73 & 0.028 & -2.654 & 0.027 \\ 
    \addlinespace

    \hline
    \end{tabular}
    \begin{tablenotes}
         \item Note. (1) Star cluster name. (2) R.A. in degrees (J2000). (3) Decl. in degrees (J2000). (4) Galactic reddening $E(B-V)$ in mag. (5) Angular radius of the cluster in degrees from MWSC. (6) Distance modulus of the cluster in mag. (7) Metallicity of the cluster. (8) Logarithm of the average age of the stars in the cluster in years.
         \item [a] Proper motion (in mas yr$^{-1}$) and the number of member stars derived from our catalog.
         \item [b] Clusters proper motion from \citet{vasiliev2021gaia}.
    \end{tablenotes}
    \label{tab:cluster}
    \end{threeparttable}
    }
\end{sidewaystable}

\end{document}